\documentclass[a4paper,11pt]{article}
\usepackage{color}
\pdfoutput=1 

\usepackage{jheppub} 

\usepackage[T1]{fontenc} 
\usepackage{booktabs} 
\usepackage{multirow}

\usepackage[hang,small,bf]{caption}
\usepackage[subrefformat=parens]{subcaption}
\usepackage{tikz}
\usepackage{pgfplots} 
\usetikzlibrary{calc}
\usetikzlibrary{datavisualization}
\usetikzlibrary{datavisualization.formats.functions}
\usetikzlibrary{backgrounds}
\usetikzlibrary{patterns}
\usetikzlibrary{decorations.pathreplacing}
\usetikzlibrary{decorations.text}
\usetikzlibrary{decorations.markings}
\usetikzlibrary{decorations.pathmorphing}
\usetikzlibrary{trees}
\usetikzlibrary{quotes,angles}
\usetikzlibrary{shapes, calc, intersections, arrows, plotmarks, decorations, tikzmark, angles, quotes, babel, positioning}
\usetikzlibrary{arrows.meta}

\tikzstyle{vecArrow} = [thick, decoration={markings,mark=at position
   1 with {\arrow[semithick]{open triangle 60}}},
   double distance=1.4pt, shorten >= 5.5pt,
   preaction = {decorate},
   postaction = {draw,line width=1.4pt, white,shorten >= 4.5pt}]
\tikzstyle{innerWhite} = [semithick, white,line width=1.4pt, shorten >= 4.5pt]

\tikzset{
particle/.style={draw=black, postaction={decorate},
decoration={markings,mark=at position .5 with {\arrow[black]{stealth}}}},
antiparticle/.style={draw=black, postaction={decorate},
decoration={markings,mark=at position .5 with {\arrow[black]{stealth reversed}}}},
sfermion/.style={draw=black, dashed, postaction={decorate},
decoration={markings,mark=at position .5 with {\arrow[black]{stealth}}}},
antisfermion/.style={draw=black, dashed, postaction={decorate},
decoration={markings,mark=at position .5 with {\arrow[black]{stealth reversed}}}},
gluon/.style={decorate, draw=black,
    decoration={coil,amplitude=4pt, segment length=4pt}},
photon/.style={decorate, decoration={snake}}
}

\renewcommand{\eqref}[1]{Eq.~\ref{#1}}
\newcommand{\figref}[1]{Figure~\ref{#1}}
\newcommand{\tabref}[1]{Table~\ref{#1}}
\newcommand{\secref}[1]{Section~\ref{#1}}
\newcommand{\appref}[1]{Appendix~\ref{#1}}
\newcommand{\mathrmbf}[1]{\mathrm{\mathbf{#1}}}
\newcommand{\I}{\mathrm{I}}
\newcommand{\II}{\mathrm{I\hspace{-1.2pt}I}}

\newcommand{\fpr}{\mathrm{FPR}}

\newcommand{\ParT}{\mathrm{ParT}}
\newcommand{\AM}{\mathrm{AM}}
\newcommand{\PT}{\mathrm{P_T}}
\newcommand{\PQ}{\mathrm{P_Q}}
\newcommand{\HT}{\mathrm{H_T}}
\newcommand{\HQ}{\mathrm{H_Q}}
\newcommand{\VT}{\mathrm{V_T}}
\newcommand{\VQ}{\mathrm{V_Q}}
\newcommand{\MF}{\mathrm{MF}}
\newcommand{\kin}{\mathrm{kin}}
\newcommand{\rcount}{\mathrm{count}}
\newcommand{\subj}{\mathrm{subj}}

\title{Jet Classification Using High-Level Features from Anatomy of Top Jets}
\author[a,b,1]{Amon Furuichi,\note{on leave to Sokendai. }}
\author[c]{Sung Hak Lim,}
\author[a, d]{Mihoko M. Nojiri}

\affiliation[a]{Graduate University for Advanced Studies (SOKENDAI) \\ 
Oho 1-1, Tsukuba, Ibaraki 305-0801, Japan}
\affiliation[b]{Department of Physics, Nagoya University \\
Furo-cho, Chikusa-ku, Nagoya, Aichi 464-8602 Japan}
\affiliation[c]{NHETC, Department of Physics and Astronomy, Rutgers, The State University of New Jersey, \\
136 Frelinghuysen Road, Piscataway, New Jersey 08854, USA}
\affiliation[d]{Theory Center, IPNS, KEK, \\
Oho 1-1, Tsukuba, Ibaraki 305-0801, Japan}

\emailAdd{amon@post.kek.jp}
\emailAdd{sunghak.lim@rutgers.edu}
\emailAdd{nojiri@post.kek.jp}

\abstract{
Recent advancements in deep learning models have significantly enhanced jet classification performance by analyzing low-level features (LLFs).
However, this approach often leads to less interpretable models, emphasizing the need to understand the decision-making process and to identify the high-level features (HLFs) crucial for explaining jet classification.
To address this, we consider the top jet tagging problems and introduce an analysis model (AM) that analyzes selected HLFs designed to capture important features of top jets.
Our AM mainly consists of the following three modules: a relation network analyzing two-point energy correlations, mathematical morphology and Minkowski functionals for generalizing jet constituent multiplicities, and a recursive neural network analyzing subjet constituent multiplicity to enhance sensitivity to subjet color charges.
We demonstrate that our AM achieves performance comparable to the Particle Transformer (ParT) while requiring fewer computational resources in a comparison of top jet tagging using jets simulated at the hadronic calorimeter angular resolution scale. 
Furthermore, as a more constrained architecture than ParT, the AM exhibits smaller training uncertainties because of the bias-variance tradeoff.
We also compare the information content of AM and ParT by decorrelating the features already learned by AM.
Lastly, we briefly comment on the results of AM with finer angular resolution inputs.

}

\pgfplotsset{compat=1.18}
\begin{document}
\maketitle
\flushbottom


\section{Introduction}

Understanding how to identify the initiating particle of jets in hadron collisions has been a crucial challenge in pursuing physics beyond the Standard Model.
This problem becomes increasingly important as the Large Hadron Collider (LHC) accumulates substantial events for discovering new physics signatures at energies above the electroweak scale.
The search involves the identification of the boosted heavy particles, such as $W/Z$ bosons, Higgs bosons, and top quarks, produced in hard processes with high energies. 
Those boosted particles often create collimated clusters similar to the jets originating from light quarks and gluons (QCD jets).
Therefore, designing and understanding taggers separating the heavy particle-initiated jets from the QCD jets has been a crucial problem of collider physics, essential for maximizing the discovery potential of the LHC and future colliders.

Modern methods for jet tagging are based on deep learning algorithms to maximize the utilization of information in a jet.
Deep learning classifiers are based on artificial neural networks capable of handling high-dimensional data and correlation among them. 
The flexibility and capability allow us to build a tagger directly analyzing low-level features (LLFs) like jet constituents.
The early studies have concentrated on network architectures that analyze jets in representations close to those popular in computer science, such as jet images \cite{Cogan:2014oua, Almeida:2015jua, deOliveira:2015xxd, Komiske:2016rsd, Kasieczka:2017nvn, Macaluso:2018tck, Farina:2018fyg, Lin:2018cin} and clustering sequences \cite{Guest:2016iqz, Louppe:2017ipp, Cheng:2017rdo}.
These deep learning-based classifiers have significantly outperformed traditional methods, which rely on a limited number of high-level features (HLFs) designed to capture specific substructural features of heavy particle-initiated jets \cite{Butterworth:2008iy, Kaplan:2008ie, Plehn:2010st, Plehn:2011sj, Thaler:2010tr, Thaler:2011gf, Larkoski:2013eya, Larkoski:2014gra} by orders of magnitude. 
Consequently, deep learning has been regarded as a promising approach for boosting the sensitivity of the LHC experiments in detecting new physics.

The state-of-the-art (SoTA) models in classification performances are based on graph neural networks \cite{graphnn} and transformers \cite{NIPS2017_3f5ee243}.
First examples in jet physics include ParticleNet \cite{Qu:2019gqs} based on dynamic graph convolutional neural networks \cite{10.1145/3326362} and Particle Flow Network \cite{Komiske:2018cqr} based on deep sets \cite{NIPS2017_f22e4747}. 
ParticleNet outperformed image-based approaches in top jet tagging \cite{Kasieczka:2019dbj} and has gained significant attention \cite{CMS-PAS-BTV-22-001}.
Graph neural networks are also convenient in fusing physics constraints, such as imposing IRC-safety in terms of energy correlators \cite{NIPS2017_f22e4747, Chakraborty:2020yfc}, using symmetry equivariant layers \cite{Gong:2022lye, Bogatskiy:2022czk} and using Lund plane \cite{Dreyer:2018nbf, Dreyer:2020brq} inspired correlations between jet constituents \cite{Qu:2022mxj}.
The most advanced taggers currently are graph neural networks that incorporate physics knowledge: Particle Transformer \cite{Qu:2022mxj}, LorentzNet \cite{Gong:2022lye}, and PELICAN \cite{Bogatskiy:2022czk}, as the additional constraint simplifies the problem while preserving networks' capabilities.

Deep learning-based taggers have shown impressive performance in learning all the detailed patterns of given jets; however, there are concerns about systematic uncertainties, particularly regarding the learning of features from soft QCD.
Since the dynamics of the strong interaction at low energies are not perturbatively described, simulations rely on phenomenological models tuned to experimental data instead \cite{Buckley:2011ms, Bierlich:2022pfr, Bellm:2015jjp, Sherpa:2019gpd}.
Consequently, the soft features in simulated jets can vary significantly from one simulation to another.
This systematic variance poses a challenge in building taggers, especially the supervised classifiers trained on simulated jets, as they may become sensitive to the choice of simulation setups \cite{Barnard:2016qma, Chakraborty:2019imr, Chakraborty:2020yfc, Lim:2020igi, Dreyer:2020brq, Ghosh:2021hrh}, though in some cases, the simulation dependencies are minor \cite{Cheung:2022dil}.
A thorough understanding of the physics information utilized by each tagger is crucial to assess and minimize the systematic uncertainties and simulation bias of taggers.

To address this problem of understanding deep learning-based jet taggers, we revisit the classifiers based on HLFs.
Compared to classifiers analyzing LLFs, taggers focusing on HLFs are more constrained and offer clearer interpretations in terms of physics.
These characteristics make HLF-based taggers more interpretable, particularly when analyzing how HLFs influence network predictions \cite{Komiske:2017aww, Komiske:2018cqr, Chakraborty:2019imr}.
Additionally, constrained taggers typically exhibit smaller training uncertainties, such as variation in network predictions due to random initialization of networks or train/validation dataset splitting \cite{Chakraborty:2020yfc, Das:2022cjl}.
This phenomenon is known as the bias-variance tradeoff, which suggests that constrained models tend to have less variance than more general models at the cost of performance.

What if HLF-based taggers are competitive with more general taggers such as graph neural networks?
There would be two immediate advantages: network predictions would have less training uncertainty, and simpler taggers would require fewer computational resources. 
In principle, if we could design a set of HLFs that fully capture features crucial for jet tagging, we could develop a high-performing jet tagger based on HLFs. Inputs used by such taggers highlight the distribution that should be calibrated by the data or give us the clear benchmarks that event simulation should satisfy.

This paper is structured as follows to address the challenges discussed above.
\secref{subsec:AMandHL} introduces our analysis model, which consists of modules analyzing groups of HLFs that capture specific characteristic features of top jets.
This model is an extension of our previous works \cite{Lim:2018toa, Chakraborty:2019imr, Chakraborty:2020yfc, Lim:2020igi}. 
Specifically, we will revisit a set of constituent multiplicities sensitive to subjet color charges, which are also important features in top jet tagging.
We will introduce a recursive neural network analyzing these numbers to our framework.

In \secref{sec:compare}, we compare the performance of the analysis model to one of the state-of-the-art jet taggers, Particle Transformer (ParT) \cite{Qu:2022mxj}.
As a benchmark, we focus on top jet classification problems using various simulations, including \texttt{Pythia} \cite{Bierlich:2022pfr}, \texttt{Herwig} \cite{Bellm:2015jjp}, and \texttt{Vincia} \cite{Ritzmann:2012ca, Fischer:2016vfv}.
Additionally, we compare jets simulated by different generators to assess each network's ability to differentiate simulations.
For simplicity, our discussion will concentrate on features at the angular scale above the hadronic calorimeter scale, 0.1.
We will demonstrate that our model has a classification performance comparable to that of ParT, emphasizing that all the introduced HLFs are crucial for achieving this level of performance.

In \secref{sec:ex.margin}, we estimate the statistical significance of the difference between our analysis model and ParT using the bootstrap method. 
Our model exhibits smaller training uncertainties while maintaining performance comparable to ParT, within the range of $2\sim3\sigma$.
We also assess the difference in information learned by the two models by penalizing information in ParT that has already been learned by the analysis model.

\secref{sec:ecal_scale} discusses the potential high-level features that start appearing below the hadronic calorimeter scale and presents the classification performance of AM tweaked for the electromagnetic calorimeter scale analysis without accounting for the additional high-level features. We summarize our findings and conclude in \secref{sec:conclusion}.


\section{Analysis Model Using High-Level Features }\label{subsec:AMandHL}

\subsection{Anatomy of Top Jets and Feature-wise Analysis Modules}

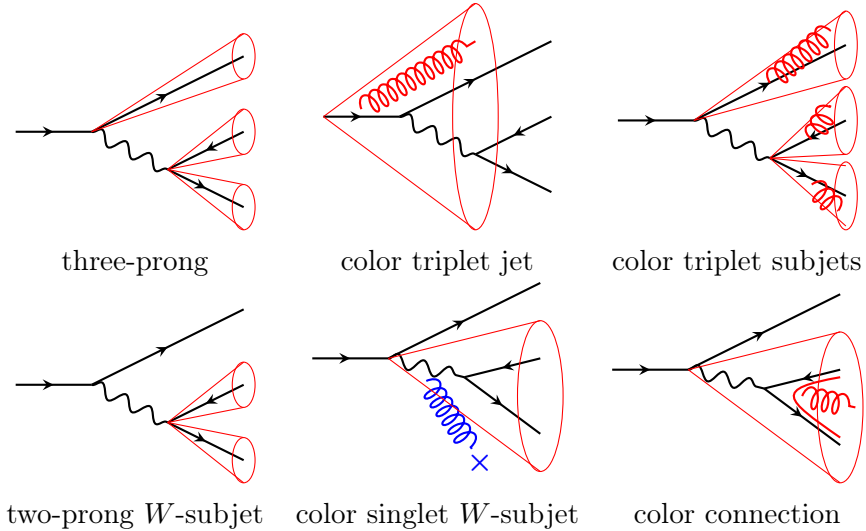
\begin{figure}
    \centering
    \begin{tabular}{ccc}
         \begin{tikzpicture}
             \draw [particle, thick] (0,0) -- (1.0,0);
             \draw [particle, thick] (1,0) -- (3.0,1.);
             \draw [photon, thick] (1,0) -- (2.0,-0.5);
             \draw [particle, thick] (3,0) -- (2.0,-0.5);
             \draw [particle, thick] (2.0,-0.5) -- (3.0,-1);
             \draw [red] (1.,0.) -- (3,1+0.3);
             \draw [red] (1.,0.) -- (3,1-0.3);
             \draw [red] (3,1) ellipse (0.15 and 0.3);
             \draw [red] (2.,-0.5) -- (3,0+0.3);
             \draw [red] (2.,-0.5) -- (3,0-0.3);
             \draw [red] (3,0.0) ellipse (0.15 and 0.3);
             \draw [red] (2.,-0.5) -- (3,-1+0.3);
             \draw [red] (2.,-0.5) -- (3,-1-0.3);
             \draw [red] (3,-1.0) ellipse (0.15 and 0.3);
         \end{tikzpicture}
         &
         \begin{tikzpicture}
             \draw [particle, thick] (0,0) -- (1.0,0);
             \draw [particle, thick] (1,0) -- (3.0,1.);
             \draw [photon, thick] (1,0) -- (2.0,-0.5);
             \draw [particle, thick] (3,0) -- (2.0,-0.5);
             \draw [particle, thick] (2.0,-0.5) -- (3.0,-1);
             \draw [gluon, thick, red] (0.5,0.1) -- (2.0,1.);
             \draw [red] (0.,0.) -- (2,0+1.5);
             \draw [red] (0.,0.) -- (2,0-1.5);
             \draw [red] (2,0) ellipse (0.3 and 1.5);
         \end{tikzpicture}
         &
         \begin{tikzpicture}
             \draw [particle, thick] (0,0) -- (1.0,0);
             \draw [particle, thick] (1,0) -- (3.0,1.);
             \draw [photon, thick] (1,0) -- (2.0,-0.5);
             \draw [particle, thick] (3,0) -- (2.0,-0.5);
             \draw [particle, thick] (2.0,-0.5) -- (3.0,-1);
             \draw [gluon, thick, red] (2.0,0.5) -- (2.8,1.2);
             \draw [red] (1.,0.) -- (3,1+0.45);
             \draw [red] (1.,0.) -- (3,1-0.45);
             \draw [red] (3,1) ellipse (0.15 and 0.45);
             \draw [gluon, thick, red] (2.55,-0.25) -- (2.8,0.2);
             \draw [red] (2.,-0.5) -- (3,0+0.45);
             \draw [red] (2.,-0.5) -- (3,0-0.45);
             \draw [red] (3,0.0) ellipse (0.15 and 0.45);
             \draw [gluon, thick, red] (2.55,-0.85) -- (2.9,-1.2);
             \draw [red] (2.,-0.5) -- (3,-1+0.4);
             \draw [red] (2.,-0.5) -- (3,-1-0.4);
             \draw [red] (3,-1.0) ellipse (0.15 and 0.45);
         \end{tikzpicture}
         \\
         three-prong & color triplet jet & color triplet subjets
         \\
         \begin{tikzpicture}
             \draw [particle, thick] (0,0) -- (1.0,0);
             \draw [particle, thick] (1,0) -- (3.0,1.);
             \draw [photon, thick] (1,0) -- (2.0,-0.5);
             \draw [particle, thick] (3,0) -- (2.0,-0.5);
             \draw [particle, thick] (2.0,-0.5) -- (3.0,-1);
             \draw [red] (2.,-0.5) -- (3,0+0.3);
             \draw [red] (2.,-0.5) -- (3,0-0.3);
             \draw [red] (3,0.0) ellipse (0.15 and 0.3);
             \draw [red] (2.,-0.5) -- (3,-1+0.3);
             \draw [red] (2.,-0.5) -- (3,-1-0.3);
             \draw [red] (3,-1.0) ellipse (0.15 and 0.3);
         \end{tikzpicture}
         & 
         \begin{tikzpicture}
             \draw [particle, thick] (0,0) -- (1.0,0);
             \draw [particle, thick] (1,0) -- (3.0,1.);
             \draw [photon, thick] (1,0) -- (2.0,-0.25);
             \draw [particle, thick] (3,0) -- (2.0,-0.25);
             \draw [particle, thick] (2.0,-0.25) -- (3.0,-1);
             \draw [gluon, thick, blue] (1.5,-0.3) -- (2.1,-1.20);
             \node [draw=none, anchor=north west] at (1.9,-1.1) {\color{blue}$\boldsymbol{\times}$};
             \draw [red] (1.,0.) -- (3,-0.5+1.0);
             \draw [red] (1.,0.) -- (3,-0.5-1.0);
             \draw [red] (3,-0.5) ellipse (0.3 and 1.0);
         \end{tikzpicture}
         &
         \begin{tikzpicture}
             \draw [particle, thick] (0,0) -- (1.0,0);
             \draw [particle, thick] (1,0) -- (3.0,1.);
             \draw [photon, thick] (1,0) -- (2.0,-0.25);
             \draw [particle, thick] (3,0) -- (2.0,-0.25);
             \draw [particle, thick] (2.0,-0.25) -- (3.0,-1);
             \draw[-, thick, red] (3.0,-0.9)
    to[out=145,in=190,distance=0.9cm] (3.0,-0.1);
             \draw [gluon, thick, red] (2.5,-0.35) -- (3.2,-0.5);
             \draw [red] (1.,0.) -- (3,-0.5+1.0);
             \draw [red] (1.,0.) -- (3,-0.5-1.0);
             \draw [red] (3,-0.5) ellipse (0.3 and 1.0);
         \end{tikzpicture}
         \\
         two-prong $W$-subjet & color singlet $W$-subjet & color connection
    \end{tabular}
    \caption{Illustrations of characteristic features of top jets.  The three diagrams at the bottom are the $W$-subjet features.
    Blue-crossed radiation in the color singlet $W$-subjet illustration denotes suppressed radiation.}
    \label{fig:top_anatomy}
\end{figure}

We design a neural network analyzing selected HLFs, capturing the most significant characteristic features of top jets, listed in \figref{fig:top_anatomy}.
Firstly, the top quark in a top jet decays into three quarks through the process $t \rightarrow b W \rightarrow b q \bar{q} $, resulting in a jet with a distinctive three-prong substructure.
Among these three prongs, two subjets originate from the decay of W-boson, and they should satisfy the W-boson decay kinematics.

In addition to these topological substructures, color substructures appear in various scales.
Since the entire system originates from a quark, the top jet exhibits a color triplet radiation pattern at the jet radius scale.
We observe two important color features at the subjet level, where three prongs are visible.
The W-boson subjet comes from a color singlet, resulting in confined radiation \cite{Larkoski:2023xam}, and a color connection between the two prongs is expected \cite{Gallicchio:2010sw}.
The presence of the three leading quark subjets is also a distinctive feature setting top jets apart from QCD jets, which consist of subjets from both quarks and gluons \cite{Pumplin:1992bv}. 
Therefore, careful consideration of these color substructures is crucial in the analysis.

Note that we do not explicitly consider the color connections that extend beyond the jet boundary in this paper.
While these external color connections are still features of top jets and can enhance the tagging performance in a supervised setup,
they are highly process-dependent, potentially making taggers less process-agnostic \cite{Lee:2023tfx}.
Although the SoTA models are generally able to exploit this information to achieve the best tagging performance, our model will not directly address the external color connections. 
Instead, we will leave our model to use the information encoded in the selected HLFs and evaluate how effectively our setup can match the tagging performance of SoTA models.

For analyzing the selected HLFs derived from top jet anatomy, we introduce our AM, consisting of neural networks that take HLFs designed to capture top jets' characteristic features fully.
The AM consists of analysis modules $\Phi^A$, which analyze a set of HLFs $x_A$, where $A$ is an HLF label. 
We denote the network output vector as $z_A$, i.e.,
\begin{equation}
    z_A = \Phi^A (x_A, x_{\kin}),
\end{equation}
where $x_{\kin}$ is a set of global features provided to all the analysis modules.
The outputs $z_A$ are then converted to a single output $\hat{y}$ by an MLP $\Phi$,
\begin{equation}
    \hat{y} = \Phi \left( \bigcup_A \, z_A  \right),
\end{equation}
and it will be trained by minimizing the binary cross-entropy loss function. 
\footnote{For training, we use the same number of data sets to make the prior of each class to be 0.5.}
\begin{figure}[t]
    \centering
    \includegraphics[width=0.5\textwidth]{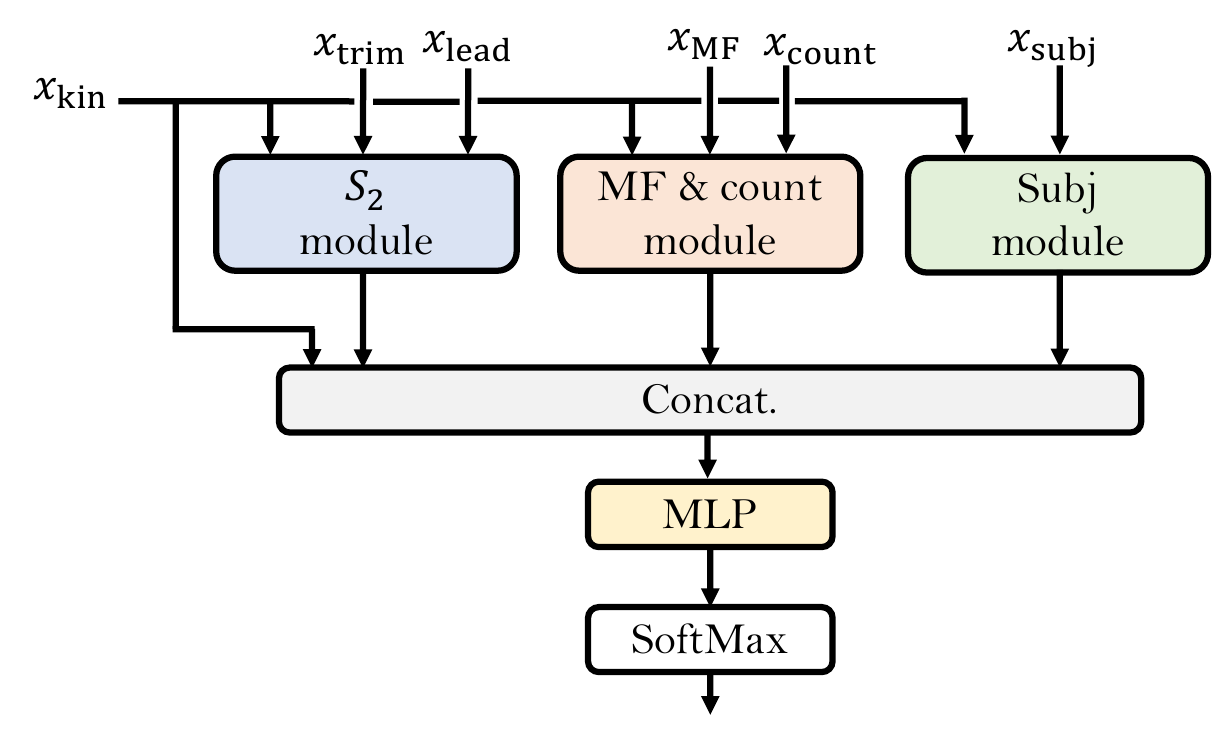}
    \caption{A schematic diagram of our AM combining $x_{\kin}$
    (\secref{subsec:pII}), $x_{S_2}=x_\mathrm{trim},x_\mathrm{lead}$ (\secref{subsec:pIII}),  $x_\rcount$ and $x_\MF$
     (\secref{subsec:pIV}), and $x_\subj$ (\secref{subsec:pV}) for the inference. }
    \label{fig:all}
\end{figure}

In the following subsections, we describe each of the HLFs and their analysis modules. 
\figref{fig:all} shows a schematic diagram of the AM combining the HLFs motivated from the following.
\begin{itemize}
    \item (\secref{subsec:pII})
    jet kinematics, $x_{\kin}$.
    \item (\secref{subsec:pIII})
    IRC-safe energy correlations, $x_{S_2}=\{ x_\mathrm{trim},x_\mathrm{lead} \}$ \cite{Lim:2018toa, Chakraborty:2019imr, Chakraborty:2020yfc}.
    \item (\secref{subsec:pIV})
    jet color charge: $p_T$-scale conditioned jet constituent multiplicity distribution $x_\rcount$ and its generalization $x_\MF$ \cite{Chakraborty:2020yfc, Lim:2020igi}.
    \item (\secref{subsec:pV})
     subjet momenta $x_{\subj}$ and its constituent multiplicity for its color charge $x_\subj^{ex}$.
\end{itemize}

The $p_T$-conditioned constituent multiplicities $x_\rcount$ and the last subjet analysis module are the main new features of this paper.
In \secref{sec:compare}, we will show that the jet classifier built from AM performs nearly equal to ParT.
Namely, the selected HLFs are sufficient for describing the top jet classifier based on ParT.
This setup further allows us to determine which features are more relevant in the classification by comparing tagging performance when some HLFs are dropped.

\subsection{Global Features: Basic Jet Kinematics}\label{subsec:pII}
We first select the global features $x_\kin$ mainly to provide all the analysis modules with the characteristic angular scales of jets, which are given in terms of the jet mass $m$ and transverse momentum $p_T$.
In particular, we use the kinematical variables of the whole jet $\mathrmbf{J}$ and the trimmed jet $\mathrmbf{J}_\mathrm{trim}$ to capture the angular scale of top quark decay.\footnote{For jet trimming, we use $k_T$ algorithm \cite{Catani:1993hr, Ellis:1993tq}with radius 0.2 and keep subjets whose energy fraction is larger than 0.05.}
We additionally provide the variables of the leading $p_T$ subjet  $\mathrmbf{J}_\mathrm{lead}$ to implicitly capture the W-boson decay scale.\footnote{The leading $p_T$ subjet is the highest $p_T$ anti-$k_T$ subjet \cite{Cacciari:2008gp} with radius 0.2.}
The resulting global features consist of the following inputs:
\begin{align}
x_\kin=(p_{T,\mathrmbf{J}}, m_\mathrmbf{J}, p_{T,\mathrmbf{J}_\mathrm{trim}}, m_{\mathrmbf{J}_\mathrm{trim}}, p_{T,\mathrmbf{J}_\mathrm{lead}}, m_{\mathrmbf{J}_\mathrm{lead}}). 
\end{align}

\subsection{IRC-Safe Energy Correlators: Relation Network and Two-Point Energy Correlation Spectrum}\label{subsec:pIII}

The topological features and color charge characteristics are often captured by IRC safe energy correlators \cite{Tkachov:1995kk}, for example, $N$-subjettiness \cite{Thaler:2010tr} and jet width/girth \cite{Gallicchio:2011xq}.
To build an analysis module covering the topological and color charge features described by the energy correlators, we use a relation network (RN) with IRC safety constraint \cite{Lim:2018toa}.

The vanilla RN \cite{DBLP:journals/corr/RaposoSBPLB17, NIPS2017_7082} is a fully connected graph neural network that only uses edge features between vertices.
Since perturbative phenomena in jets, such as particle decays and parton showers, are often described by two-body kinematics, an RN is a good choice for the first feature extraction.
The RN consists of two networks: $\Phi_E$ for modeling derived edge features between two vertices and $\Phi_G$ for mapping accumulated edge features to the global output feature.
The network output is then written as follows:
\begin{equation}
\Phi^G \left[ \sum_{i \in a, j \in b} \Phi^E(v_i, v_j) \right],
\end{equation}
where $v_i$ and $v_j$ are the vertex features, and the sum runs over all the vertices representing jet constituents in sets $a$ and $b$.

With IRC safety and additional translation and rotation invariances on the $(\eta, \phi)$ plane, the summand must be a two-point energy correlator with angular weighting function $W(R)$ that solely depends on the relative distance between two constituents $R_{ij} = \sqrt{(\eta_i-\eta_j)^2+ ( \phi_i-\phi_j)^2}$, i.e.,
\begin{equation}
\sum_{i \in a, j \in b}  \Phi^E(v_i, v_j) = \sum_{i \in a, j \in b}  p_{T,i} p_{T,j} W(R_{ij}).
\end{equation}

This RN can be significantly simplified by introducing an IRC safe \emph{two-point correlation spectrum} \cite{Lim:2020igi,Chakraborty:2020yfc,Lim:2018toa,Chakraborty:2019imr},
\begin{equation}
  S_{2,ab}(R)
  =
  \sum_{i \in a, j \in b}  p_{T,i} p_{T,j} \delta(R - R_{ij}).
\label{eq:s2}
\end{equation}
The term $S_{2,ab}(R) \, dR$ can be interpreted as a $p_{T,i} p_{T,j}$ weighted histogram of the angular distance of jet constituent pairs.
The double summations in edge feature accumulation are then simplified to a single integral,
\begin{equation}
\sum_{i \in a, j \in b}  \Phi^E(v_i, v_j) = \int dR \; S_{2,ab}(R)  W(R).
\label{eqn:RN_2pt_model}
\end{equation}
Note that this expression covers two-point energy flow polynomials $\mathrm{EFP}^n_2$ \cite{Komiske:2017aww,Komiske:2018cqr} when the angular weighting function $W(R)$ is a polynomial. 
We evaluate the integral with a simple rectangle rule with finite bin size $\Delta R = 0.1$; the integral turns into a linear combination of binned two-point energy correlation spectrum $S_{2,ab}(R;\Delta R)= \int_{R}^{R+\Delta R} dR \, S_{2,ab}(R)$.
The linear combination coefficients can then be absorbed into the weight matrix of the first linear layer in $\Phi_G$.
The RN simplifies an MLP by taking $S_{2,ab}(R;\Delta R)$'s as inputs.

\begin{figure}[ht]
 \begin{minipage}[ht]{0.45\linewidth}
  \centering
  \includegraphics[keepaspectratio,width=0.5\textwidth]{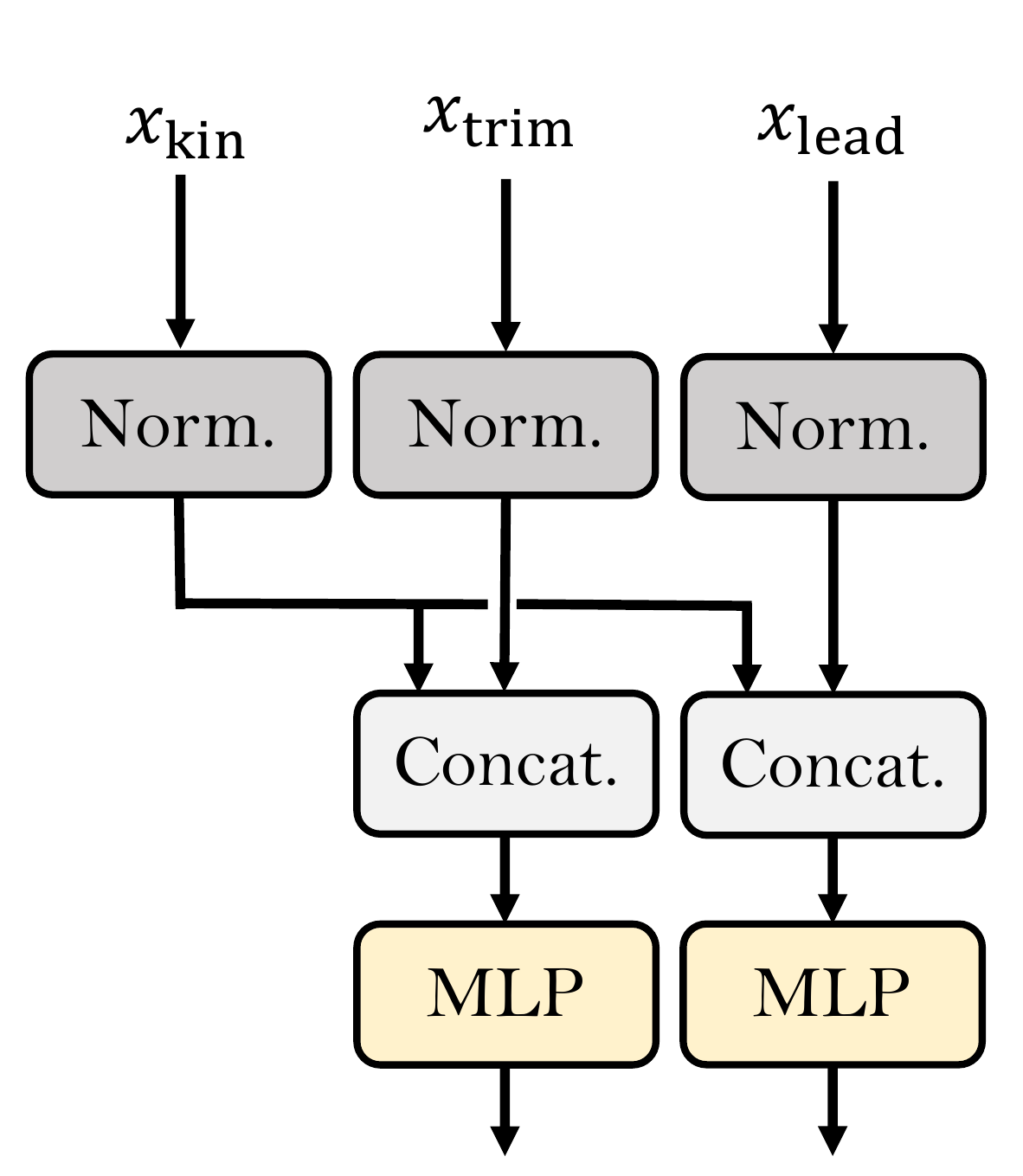}
  \subcaption{}
  \label{fig:s2_module}
 \end{minipage}
 \begin{minipage}[ht]{0.45\linewidth}
  \centering
  \includegraphics[keepaspectratio,width=1\textwidth]
  {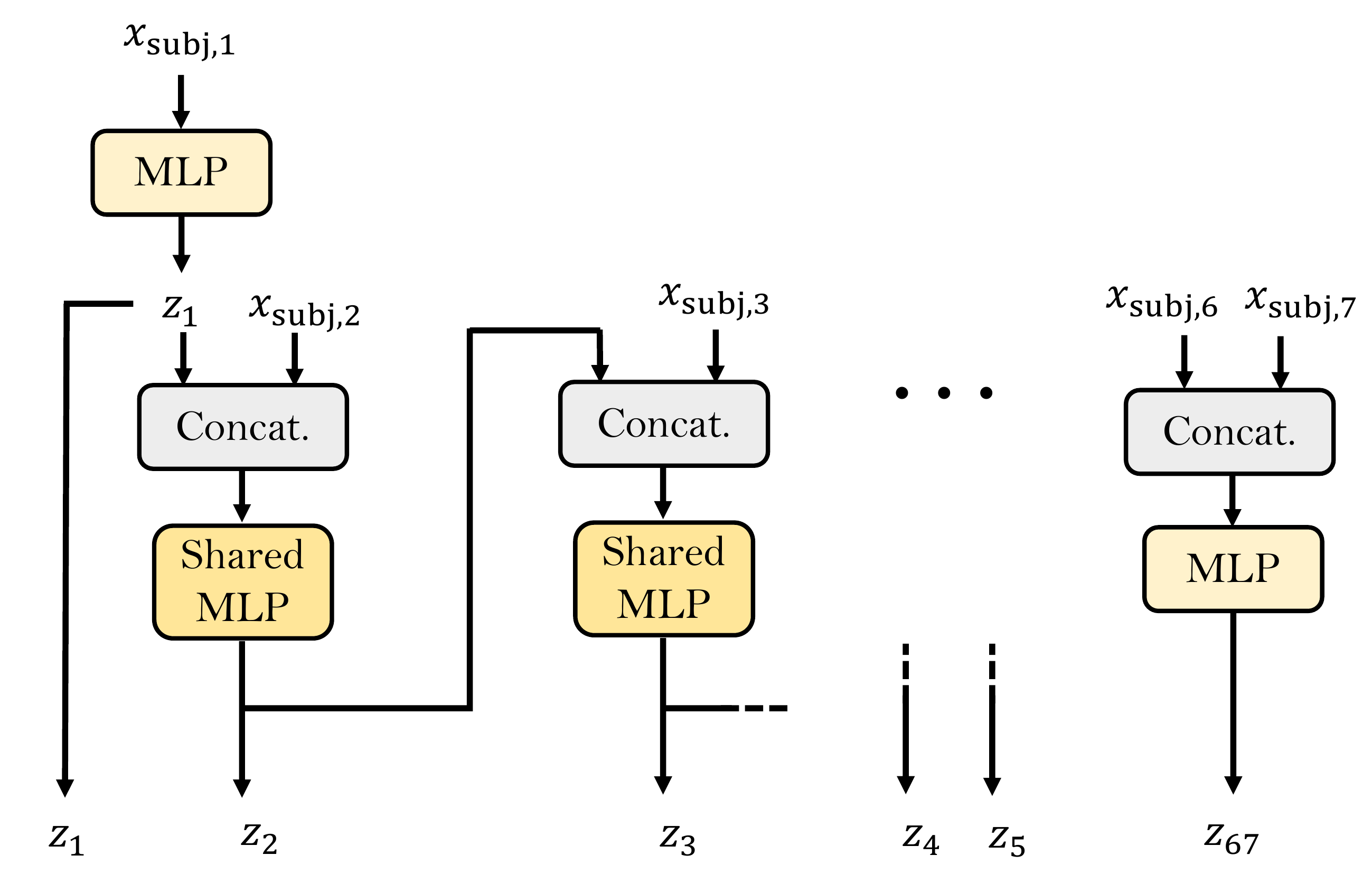}
  \subcaption{}
  \label{fig:recursive}
 \end{minipage}
 \caption{(a) A schematic diagram of relation network. (b) A schematic diagram of recurrent neural network analyzing subjet constituent multiplicities and subjet kinematics.}
\end{figure}

We again employ the trimmed jet $\mathrmbf{J}_{\mathrm{trim}}$ and leading $p_T$ subjet $\mathrmbf{J}_{\mathrm{lead}}$ for the constituent subsets to explicitly capture correlations in hard vs.~soft radiations and the leading jet vs.~others, respectively. 
In the $\mathrmbf{J}_{\mathrm{trim}}$ analysis module, we use $a,b\in\{\mathrmbf{J}_\mathrm{trim}, \mathrmbf{J}_\mathrm{trim}^c = \mathrmbf{J}-\mathrmbf{J}_\mathrm{trim}\}$ and denote the set of binned $S_{2,ab}$'s as $x_\mathrm{trim}$.
Similarly in the $\mathrmbf{J}_{\mathrm{lead}}$ analysis module, we use $a,b\in\{\mathrmbf{J}_\mathrm{lead}, \mathrmbf{J}^c_\mathrm{lead} = \mathrmbf{J}-\mathrmbf{J}_\mathrm{lead}\}$ and refer the set of binned $S_{2,ab}$'s as $x_\mathrm{lead}$.
It is noteworthy that the complementary sets $\mathrmbf{J}_\mathrm{trim}^c$ and $\mathrmbf{J}^c_\mathrm{lead}$ are formally IRC safe, and they provide a different perspective on the distribution of soft particles.
We represent the output of the two analysis modules as $z_\mathrm{trim}$ and $z_\mathrm{lead}$,
\begin{eqnarray}
    z_\mathrm{trim}
    &=&
    \Phi^\mathrm{trim}(x_\mathrm{trim},x_\kin),
    \\
    z_\mathrm{lead}
    &=&
    \Phi^\mathrm{lead}
    (x_\mathrm{lead},x_\kin).
\end{eqnarray}
The RNs $\Phi^\mathrm{trim}$ and $\Phi^\mathrm{lead}$ are modeled using MLPs with two hidden layers.
The schematic structure of the networks is illustrated in \figref{fig:s2_module}.

\subsection{Constituent Multiplicity Distributions and Mathematical Morphology}
\label{subsec:pIV}

To fully capture the color-related characteristics of jets, the IRC-safe shape variables alone are not enough; we also need IRC-unsafe variables.
For instance, counting variables play a crucial role in the quark vs.~gluon jet classification problem, where the distinguishing feature is the color charge of originating parton: color triplet or octet \cite{PhysRevD.44.2025,Gallicchio:2011xq,Frye:2017yrw}.
Such counting variables are also essential in top jet tagging, as quark vs.~gluon jet classification is needed at the subjet level.
Note that the kinematics of subjets are different in top jets and QCD jets; the quark subjets of top jets tend to have similar starting $p_T$ scale, whereas those of QCD jets do not, due to the kinematics of parton shower.
This kinematic difference in subjets suggests that conditional distributions of the counting variable given jet constituent $p_T$ can be an informative variable for capturing those color substructures of subjets.

\begin{figure}[t!]
    \centering
    \includegraphics[width=0.7\linewidth]{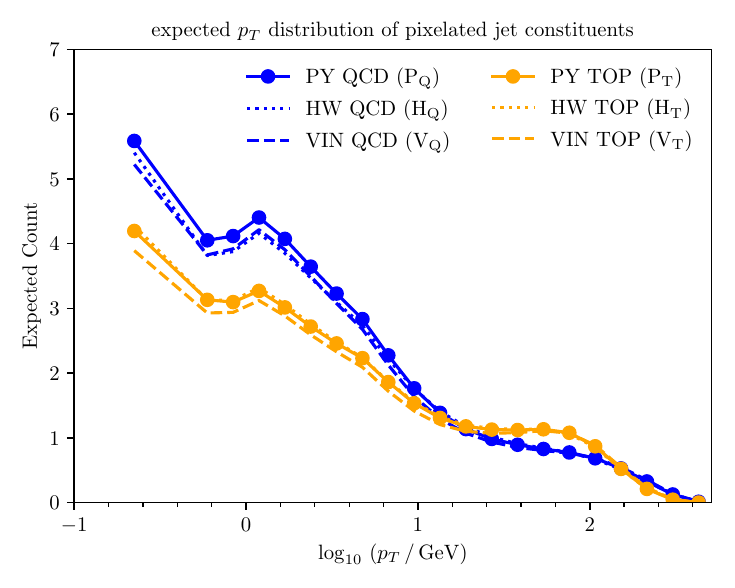}
    \caption{ Expected $p_T$ distribution of pixelated jet constituents in top and QCD jets.
    The constituents are converted into pixels on $(\eta,\phi)$ planes, with a pixel size of $0.1\times 0.1$.
    The expected counts of the histogram bins are shown, with boundaries at $p_T \in \{ 0.1, \, 0.5\times(\sqrt{2})^k|_{k=0,\cdots,20} \} \;\mathrm{GeV}$.  
    For our event samples and pre-processing details, see \secref{subsec:dataset}.
    }
    \label{fig:counts}
\end{figure}

\figref{fig:counts} shows the expected count histogram of jet constituent $p_T$, averaged over our training samples. 
(See \secref{subsec:dataset} for the detailed information on our datasets.) 
Results from three event generators -- \texttt{Pythia} (PY), \texttt{Herwig} (HW), and \texttt{Vincia} (VIN) -- are represented by solid, dotted, and dashed lines, respectively.

In the figure, QCD jet distributions are shown in blue lines. 
The HW and VIN expectations have similar trends, while PY produces slightly more constituents overall.
The cases of top jets are shown in yellow lines; the PY and HW expectations are similar, while the VIN curve is slightly lower than the others.
Note that the difference between event generators is more prominent in the soft region as this region is more sensitive to the parton shower, hadronization modeling, and their simulation tuning.

The differences in expectations between the QCD and top jets are more significant than the above systematic uncertainties between the simulations.
In the low $p_T$ region, QCD jets typically have more constituents than top jets due to gluon (sub)jets and the high mass selection.
On the other hand, the high $p_T$ region captures features of the hard process. 
The top jet histogram notably shows a prominent bump at $\sim 50$ GeV. 
This bump is mostly due to the subjets originating from the $W$ boson, and the hadronization of the corresponding boosted $q\bar{q}$ system.
This multiplicity distribution conditioned on $p_T$ captures the differences between top and QCD jets and the systematic differences between simulations.
Including this histogram and its generalization for the AM will significantly enhance the information encoding performance.

One can further extend the analysis of counting information by using Minkowski Functionals (MFs) in mathematical morphology.
MFs are complete linear basis functions of geometric measures with a notion of size, including the constituent multiplicity. 
If we use the MFs as inputs to an MLP, the resulting network effectively becomes a module analyzing geometric measures.
The first linear layer models geometric measures, and the subsequent layers take care of its non-linear transformation.
This method of constructing an analysis module is similar to the previously discussed simplification of RN with IRC safety constant.
For analyzing jet constituent distributions on the two-dimensional $(\eta,\phi)$ plane, we need to consider the three MFs: area $A$, perimeter length $L$, and Euler characteristic $\chi$.

\begin{figure*}[ht!]
\centering
\includegraphics[width=0.6\textwidth]{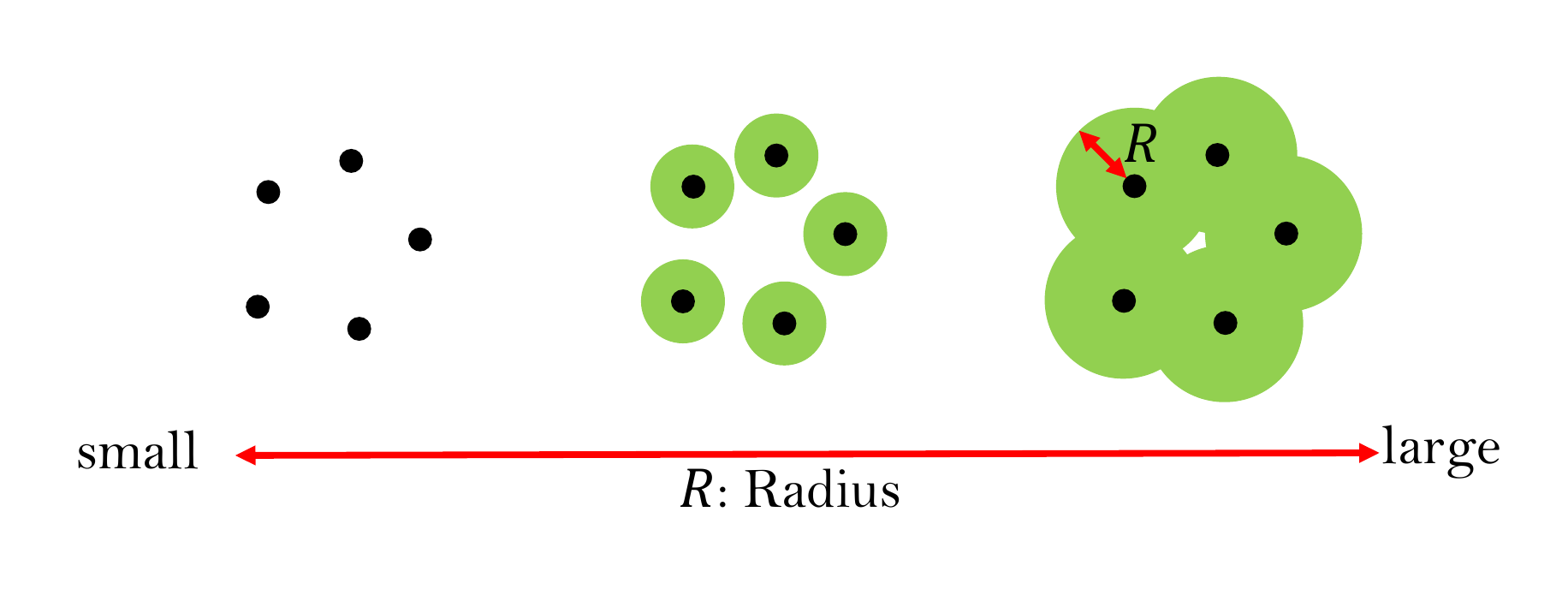}
\caption{
An illustration of the dilation of point clouds by disks. 
This dilation is used before evaluating the Minkowski Functionals.
}
\label{Fig:MF}
\end{figure*}

These MFs are used with dilation operation to encode non-trivial changes in the topology and geometry appearing at various scales, similar to persistence homology in topological data analysis \cite{10.3389/frai.2021.667963}.
Let us define a dilated image $\mathcal{I}(R)$ as a Minkowski sum of the set of constituents $\mathcal{C}$ and a disk $B(R)$ with radius $R$, i.e.,\footnote{
Note that in our previous works \cite{Chakraborty:2020yfc, Lim:2020igi}, we constructed a model using the MFs dilation by a square and compared the results with CNN—both of the models process jet images with Manhattan geometry.
In this paper, we use the disk respecting the rotational symmetry of Euclidean geometry instead.
The edge variable of ParT is invariant under rotation, and this change aligns persistence analysis through the MF module consistent with ParT. 
}
\begin{equation}
   \mathcal{I}(R) = \mathcal{C} + B(R) = \{a+b \,|\, a\in \mathcal{C}, b \in B(R)\}.
\end{equation}
We illustrate this dilation operation in \figref{Fig:MF}.
The corresponding MFs of $\mathcal{I}(R)$ are denoted as $A(R)$, $L(R)$, and $\chi(R)$.
By construction, these MFs are translation and rotation invariant.

The persistence analysis on MFs under the dilation captures changes in topology and geometry.
When the dilation is sufficiently small, the MFs of the dilated image are described by Steiner's formula, i.e., the dilated MFs are linear combinations of MFs without dilation, i.e., dots. 
For example, consider the dilation from left to middle of \figref{Fig:MF} just before the disks touch each other.
The MFs, in this case, can be expressed as follows.
\begin{equation}
A(R) = \pi R^2 \cdot \chi(0), \quad L(R) = 2 \pi R \cdot \chi(0), \quad \mathrm{and} \;\chi(R) = \chi(0). 
\label{eqn:MFs:steiner_bare}
\end{equation}
However, as soon as the disks come in contact, this relationship no longer holds, indicating that a non-trivial topological change occurred at the dilation scale.
The MFs after the dilations are computed using \texttt{shapely} package \cite{Gillies_Shapely_2023} based on \texttt{GEOS} library \cite{libgeos}.
For the numerical evaluations, the disks are approximated by regular 64-sided polygons.
The Python code used for computing the MFs is available at \url{https://github.com/sunghak-lim/mfjet}.

To make this persistence analysis conditioned on constituent $p_T$, we examine the MFs of jet constituents exceeding certain $p_T$ thresholds $p_T^{\mathrm{th}}$.
This setup yields MFs -- $A(R;p_T^{\mathrm{th}})$, $L(R;p_T^{\mathrm{th}})$, and $\chi(R;p_T^{\mathrm{th}})$ -- as functions of disk radius $R$ and the threshold $p_T^{\mathrm{th}}$.
These functions can simultaneously capture the morphology of soft constituents and hard substructures within the jet.
The MFs are sampled at the following values of $R$ and $p_T^{\mathrm{th}}$ to create a discrete set of network inputs.
\begin{eqnarray}
    \nonumber
    R 
    & \in &
   \left.
    \begin{cases}
        0.025 k + \epsilon & k= 1, \cdots, 8 \\
        0.2 + 0.05 (k-8) + \epsilon & k = 9 ,\; 10
    \end{cases}
    \right\} 
    \\
    p^\mathrm{th}_{T} 
    & \in &
    \left\{ 0,1,2,4,8  \right\} \;  \mathrm{GeV}.
    \label{eqn:MFs:pt_r_samples}
\end{eqnarray}
Here, $\epsilon$ is a small positive number to avoid the numerical instability appearing when two disks overlap only at the boundary.

Note that the diameter $2R$ is the parameter describing the correlation scale in this persistence analysis.
The first MFs with $R=0.025$ are simply the constituent multiplicities since we are using angular resolution $0.1$ and no disks are overlapping each other; the MFs will be simply given as \eqref{eqn:MFs:steiner_bare}.
For $k>1$, some disks may overlap, and MFs can capture non-trivial topological changes during the dilation.
For $R>0.2$, we switch to a coarser grid.  
Most small-scale effects are smeared out at this scale, leaving only large-scale information less sensitive to the $R$ choice, such as hard substructure and large-angle radiations. 
We limit the angular scale up to $R_\mathrm{max}=0.3$ in this paper. The dependence on the choice of $R_\mathrm{max}$ can be found in 
\secref{sec:R-dep}.

In summary, this jet morphology analysis module covering IRC-unsafe counting observables considers the following inputs:
\begin{itemize}
    \item $x_{\mathrm{count}}$: the values of jet constituent histogram bins over constituent $p_T$ within 0.1~GeV to 50~GeV.
    We use the following bin edges scaling exponentially:
    \begin{equation}
    \left[
        0.1,\;
        a_1,\; \cdots,\;
        a_{n_\mathrm{bins}-1},\;
        a_{n_\mathrm{bins}} =50
    \right]\; \mathrm{GeV},
    \quad
    a_{k} = 0.5 \cdot 10^{2 (k-1)/(n_\mathrm{bins} - 1)}\;
    \end{equation}
    We use $n_\mathrm{bins}=20$ bins for this analysis module.
    \item $x_{\mathrm{MF}}$: $A(R;p_T^{\mathrm{th}})$, $L(R;p_T^{\mathrm{th}})$, and $\chi(R;p_T^{\mathrm{th}})$ evaluated at points described in \eqref{eqn:MFs:pt_r_samples} with $R \leq 0.3$.
\end{itemize}
We simply use an MLP $\Phi^{\mathrm{MF}}$ to analyze these inputs.
The outputs $z_{\mathrm{MF}}$ are given as follows.
\begin{equation}
    z_{\mathrm{MF}} = \Phi^{\mathrm{MF}} ( x_{\mathrm{count}}, x_{\mathrm{MF}}, x_{\mathrm{kin}} ).
\end{equation}

\subsection{Recursive Neural Network for Subjet Constituent Multiplicities and Kinematics}\label{subsec:pV}

The analysis of constituent multiplicity and Minkowkski functional conditioned on $p_T$ provides information regarding the color charge of jets and its morphological behavior at each $p_T$ scale.
This approach implicitly captures the information about subjet color charges.
However, a few ambiguous cases exist, especially when two subjets have a similar $p_T$ scale. 
To address this ambiguity, we introduce an analysis module for dealing with subjet constituent multiplicity explicitly.

In our subjet analysis module, we focus on seven leading $p_T$ Cambridge-Aachen subjets \cite{Dokshitzer:1997in, Wobisch:1998wt} with radii of 0.1, 0.2, and 0.3. 
This is because the angular resolution of inputs is 0.1, and other modules handle larger angular scales. 
The choice of seven subjets is motivated by the results of $n$-subjettiness based taggers \cite{Datta:2017rhs}.
As the subjettiness analysis for top jet tagging saturates at 5-body analysis \cite{Moore:2018lsr}, considering seven subjets allows us to account for a few additional soft radiations.

We consider the following sets of features for each $k$-th subjet with radius $R$.
\begin{eqnarray}
x^\mathrm{(ex)}_{\subj,k,R} 
& = &
\{p_k, N_{c,k}\} 
\\
x_{\subj,k,R}
& = &
\{p_k\} 
\end{eqnarray}
Here, $p_k$ is the subjet $p_T$, $\eta$, $\phi$ and $m$, and $N_{c,k}$ denotes the number of constituents in the subjet.
For simplicity, we do not use the full Minkowski functionals; the $N_{c,k}$ alone provides sufficient information when this module operates together with the others.

We will analyze the above subjet features $x^\mathrm{(ex)}_{\subj,k,R}$ in a sequential setup to emphasize the subjet ordering, particularly because the first three subjets in top jets are likely to be quark jets.
For this purpose, we stack the features of $k$-th subjets with different radii as follows,
\begin{equation}
x^\mathrm{(ex)}_{\subj,k}
= 
\begin{cases}
    \{ x^\mathrm{(ex)}_{\subj,k,R} \;|\; {R=0.1,0.2,0.3} \}&
    \textrm{for} \; k\le 5
    \\
    \{ x_{\subj,k,R} \;|\; {R=0.1,0.2} \} &
    \textrm{for} \; k = 6, 7.
\end{cases}
\end{equation}
For the 6th and 7th subjets, we omit $R=0.3$ since the preceding subjets sufficiently cover most of the jets constituents, and the remaining are again minor clusters. 
We also consider $N_{c,k}$ only in the first five subjets as subsequent subjets are soft subjets, and their color charge is less relevant. 
A simpler input set $x_{\subj,k}$ without constituent multiplicities is also introduced for comparison,
\begin{equation}
x_{\subj,k} = \{ x_{\subj,k,R} \;|\; {R=0.1,0.2} \}
\     \textrm{for} \; k\le 7.
\end{equation}

To effectively accommodate the subjet features into our AM, we use a recursive neural network for the first five subjets, structured as follows:
\begin{eqnarray}
z_1
& = &
\Phi^\mathrm{1}(x^\mathrm{(ex)}_{\subj,1}, x_{\kin})
\\
z_k
& = &
\Phi^\mathrm{C}(z_{\mathrm 
k-1}, x^\mathrm{(ex)}_{\subj,k}, x_{\kin}) \quad \mathrm{for} \; k=2,\cdots,5.
\end{eqnarray}
Here, each $\Phi^\mathrm{C}$ represents a shared MLP with two hidden layers common for $k=2, \cdots, 5$. 
The 6th and 7th subjets contain less hierarchically important information; we use a separate MLP $\Phi^\mathrm{67}$ with two hidden layers to process the information.
\begin{equation}
z_{67} 
=
\Phi^\mathrm{67}(x_{\subj,6}, x_{\subj,7}, x_{\kin}).
\end{equation}
The final output sequence of this subjet analysis module, $z_{\mathrm{subj}}$, is defined as follows.
\begin{equation}
z^\mathrm{(ex)}_{\mathrm{subj}}
=
(z_1, \cdots, z_5, z_{67}),
\end{equation}
where $z^\mathrm{(ex)}_{\mathrm{subj}}$ corresponds to $x^\mathrm{(ex)}_{\subj,k}$ inputs respectively. 
\figref{fig:recursive} illustrates the schematic diagram of this network configuration. 
Note that if we use a deep set architecture \cite{NIPS2017_f22e4747} to introduce permutation symmetry for this analyzing module, the resulting network is a variant of the jet flow network \cite{Athanasakos:2023fhq} with additional subjet constituent multiplicity information.

\section{Comparing Classification Performances of Analysis Model and Particle Transformer}
\label{sec:compare}

\subsection{Datasets}\label{subsec:dataset}

In this section, we explain top jet and QCD jet datasets used for training jet classifiers.
Top jets are sampled from the hard process $pp \rightarrow t\bar{t}$, and QCD jets are from $pp \rightarrow jj$\footnote{$j$ represents either gluon or quark ($u$, $d$, $s$, $c$, or $b$).}.
Both processes are simulated at a center-of-mass energy $\sqrt{s}=13$~TeV using \texttt{MadGraph5} 3.4.2 \cite{Alwall:2014hca}.
The simulated top quarks are decayed hadronically using \texttt{madspin} \cite{Artoisenet:2012st}.
To focus on generating samples for boosted top jet classification, we exclusively generate events where the transverse momentum of top quarks in $t\bar{t}$ events and leading $p_T$ quark/gluon in dijet events is greater than 450~GeV.

For the parton shower and hadronization simulations, we use \texttt{Pythia} 8.308 \cite{Bierlich:2022pfr} and \texttt{Herwig} 7.2  \cite{Bellm:2019zci}, considering the following parton shower models,
\begin{itemize}
    \itemsep0em
    \item \texttt{Pythia}, \texttt{simple shower}, (denoted as PY)
    \item \texttt{Pythia}, \texttt{Vincia shower}, (denoted as VIN)
    \item \texttt{Herwig}, default shower, (denoted as HW)
\end{itemize}
\texttt{Vincia shower} \cite{Ritzmann:2012ca,Fischer:2016vfv} is a $p_T$-ordered parton shower model for QCD + QED/EW showers based on the antenna formalism.
This model exhibits improved color-coherence effects relative to the default \texttt{simple shower} model of \texttt{Pythia}.
We additionally include \texttt{Vincia} to assess whether our AM has enough sensitivity to distinguish different Monte Carlo simulations, particularly in PY vs.~VIN comparisons. 
Detailed results are presented in \appref{subapp:vincia}.
Regarding the simulation tune settings, the Monash tune \cite{Skands:2014pea} is used for \texttt{Pythia}, and the default tune is used for \texttt{Herwig}.
Furthermore, we simulate neutral pion decay to ensure accurate modeling of particle distribution at angular scales near 0.1.

The generated particle-level events are processed using the \texttt{Delphes} detector simulation \cite{deFavereau:2013fsa} using the default ATLAS detector card with a modification: jets are reconstructed from EFlow objects \cite{deFavereau:2013fsa, DeRoeck:2005sqi}. 
The jets are reconstructed by the anti-kt algorithm \cite{Cacciari:2008gp} with radius parameter $R=1$, which is implemented in \texttt{fastjet} \cite{Cacciari:2011ma, Cacciari:2005hq}. 
Note that the EFlow objects include information from tracks and electromagnetic calorimeters, offering a better angular resolution than hadronic calorimeter towers. 
However, to compare network architectures at the same lower angular scale threshold, we will later mask out the shorter-scale information during the construction of HLFs for AM and LLFs for ParT.

We select the leading $p_T$ jets with transverse momentum $p_{T,J} \in [500,600]$~GeV, jet mass $m_J \in [150, 200]$~GeV,
and pseudorapidity $\vert \eta \vert \le 2$ as our top jet and QCD jet samples. 
For top jets, we further require that all quarks originating from the top and W boson decay are located within a distance $R=1$ from the jet axis on the $(\eta,\phi)$-plane.
We denote the PY, VIN, and HW simulated QCD jet datasets as $\PQ$, $\VQ$, and $\HQ$, respectively.
Similarly, we use $\PT$, $\VT$, and $\HT$ to represent PY, VIN, and HW $t \bar{t}$ datasets, respectively.
The training and test dataset sizes are listed in \tabref{Table:num_events}.
While the dataset sizes are not uniform, we ensure the class balance by downsampling the larger dataset so that both classes have the same number of samples during the training.
However, during testing, we use the entire dataset without downsampling.

Note that the dataset sizes are bigger than the public top tagging challenge dataset \cite{Kasieczka:2019dbj, kasieczka_2019_2603256} in a similar kinematic range. 
The community challenge dataset, covering a slightly different $p_{T,J}$ range of $[550,650]$~GeV, does not include preselection based on jet mass.
The number of top (QCD) training dataset in the jet mass $\in [150,200]$~GeV is 476K (59K), while our QCD training datasets have about 400K samples.
Our dataset has more QCD jets in this mass range, which helps to mitigate the class imbalance during the training.

\begin{table}
\begin{center}
\begin{tabular}{lcccccc}
\toprule
&$\PT$
&$\PQ$
&$\HT$ 
&$\HQ$
&$\VT$ 
&$\VQ$\\
\midrule
Number of training data
& 524K
&384K
&395K
&418K
&403K
&329K \\ 
Number of testing data
&527K
&388K
&393K
&417K
&403K
&328K \\
\bottomrule
\end{tabular}
\caption{
The number of training and testing samples generated by \texttt{Pythia} (P), \texttt{Herwig} (H), and \texttt{Vincia} (V). 
The subscripts T and Q denote top jet and QCD jet samples, respectively. 
During the training of a binary classifier, we ensure the use of an equal number of samples for both classes by downsampling the larger dataset. Still, all available samples are used for testing. 
}
\label{Table:num_events}
\end{center}
\end{table}

\subsection{Particle Transformer}
\label{subsec:ParT}

We will mainly compare our AM results with the Particle Transformer (ParT) \cite{Qu:2022mxj}.
ParT is a transformer-based neural network \cite{NIPS2017_3f5ee243} that directly analyzes low-level features (LLFs) of jet constituents.
The ParT architecture consists of three main modules:
\begin{itemize}
    \itemsep0em
    \item 
    Particle attention blocks \cite{Qu:2022mxj}, which analyze LLFs. 
    This module is based on NormFormer architecture \cite{shleifer2021normformer} with augmented multi-head attention layers. 
    The query-key inner product in the attention layers is augmented by pairwise features inspired by LundNet \cite{Dreyer:2020brq}.
    \item 
    Class attention blocks \cite{Touvron_2021_ICCV} that transform a trainable class token \cite{dosovitskiy2021an, devlin2018bert} into an output, utilizing the final output of the particle attention blocks.
    \item
    An MLP that converts the transformed class token into classifier outputs.
\end{itemize}
The final output of ParT is then trained by minimizing the binary cross-entropy loss function.

Since our aim is to compare classifier performance with inputs having a finite angular resolution $0.1$, we will mask out small-scale information by utilizing jet images \cite{Cogan:2014oua, Almeida:2015jua, deOliveira:2015xxd}.
The jet constituents are first preprocessed as described in \cite{Lim:2020igi}.
This preprocessing involves shifting, rotating, and flipping on the $(\eta,\phi)$ coordinate to ensure that the leading subjet is at the origin, the second subjet lies on the positive $\phi$-axis ($\eta=0$ and $\phi>0$), and the third subjet is on the right side of the plane ($\eta>0$).
After the preprocessing, the jet constituents are pixelated into bins of size $\Delta \eta= \Delta \phi=0.1$. 
Pixels with non-zero energy deposits are converted into massless 4-momenta using their $(\eta,\phi)$ coordinates and energy deposits.
This set of 4-momenta is used as ParT inputs. The results using a finer resolution of $\Delta \eta=\Delta\phi= 0.025$ will be shown in \secref{sec:ecal_scale}.

As a result of the pixelation process, the inputs become relatively simpler than the original 4-momenta; therefore, we use a smaller version of ParT compared to the one presented in the original paper \cite{Qu:2022mxj}.
Given that our sample size is about 380k per class, we target a similar number of network parameters in order to maintain a positive degree of freedom.
Specifically, we halve the number of attention blocks and the width of each neural network from the default setting. 
In this configuration, ParT has 340k parameters.

\subsection{Comparing Analysis Model and Particle Transformer}

\subsubsection{Comparing Classification Performance}
\label{sec:comp:am_part}

In this subsection, we compare the classification performance between AM and ParT, specifically using samples generated by PY and HW. 
The results involving samples generated by VIN are presented in \appref{subapp:vincia}.

We particularly focus on the following four classification problems: ($\PT$~vs.~$\PQ$), ($\HT$~vs.~$\HQ$),  ($\PQ$~vs.~$\HQ$), and ($\PT$~vs.~$\HT$). 
The first two, ($\PT$~vs.~$\PQ$) and ($\HT$~vs.~$\HQ$), are the top jet tagging problems using samples generated by PY and HW, respectively. 
Comparing these two will provide insight into the generator-dependency of taggers.
The latter two, ($\PQ$~vs.~$\HQ$) and  ($\PT$~vs.~$\HT$), focus on detecting differences between jets generated by PY and HW.
These classifiers are trained to learn the likelihood ratio between two distinct Monte-Carlo simulations, which is crucial in quantifying differences between simulations.
In this paper, our initial focus is on evaluating whether the HLFs used in AM are sufficient for effectively paralleling ParT analyzing LLFs. 
In-depth analyses of systematic differences using HLFs will be future works.

\begin{table}
\begin{center}
\begin{tabular}{lcccc}
\toprule
\multirow{2}{*}{Model} & 
$\PT$ vs.~$\PQ$ & 
$\HT$ vs.~$\HQ$ & 
$\PQ$ vs.~$\HQ$
& 
$\PT$ vs.~$\HT$ 
\\
\cmidrule(lr){2-2}\cmidrule(lr){3-3}\cmidrule(lr){4-4}\cmidrule(lr){5-5}
& AUC & AUC & AUC & AUC
\\
\midrule
CNN \cite{Chakraborty:2020yfc} & 
0.940 &
0.924 &
0.577 &
0.548\phantom{0}
\\
\cmidrule[0.2pt](lr){1-5}
\small{\emph{IRC-safe AM-PIPs}} &
\\
A) $x_{\rm kin}$, $x_{S_2}$ &
0.936 &
0.919 &
0.571 &
0.569\phantom{0}
\\
B) $x_{\rm kin}$, $x_\subj$ &
0.938 &
0.922 &
0.575 &
0.578\phantom{0}
\\
C) $x_{\rm kin}$, $x_{S_2}$, $x_\subj$ &
0.941 &
0.925 &
0.579 &
0.585\phantom{0}
\\
\cmidrule[0.2pt](lr){1-5}
\small{\emph{IRC-unsafe AM-PIPs}} &
\\
D) $x_{\rm kin}$, $x_{\rm MF}$, $x_{\rm count}$ &
0.930 &
0.912 &
0.583 &
0.587\phantom{0} 
\\
E) D) + $x_{S_2}$ &
0.941 &
0.925 &
0.595 &
0.5986
\\
F) D) + $x_{S_2}$, $x_{\subj}$ &
0.942 &
0.926 &
0.597 &
0.5995
\\
Full AM&
0.943 &
0.928 &
\bf 0.600 &
0.605\phantom{0}
\\
\cmidrule[0.2pt](lr){1-5}
ParT \cite{Qu:2022mxj} & 
\bf 0.944 &
\bf 0.929 &
0.599 &
\bf 0.627\phantom{0}
\\
\bottomrule
\end{tabular}
\end{center}
\caption{
AUC for CNN, ParT, AM, and various configurations of AM-PIPs. 
The specific inputs activated in each AM-PIP are listed in the left column. 
The AUC of AM is closest to ParT in all classifications, except for $\PT$~vs.~$\HT$. 
Note that $x_\subj$ does not include subjet constituent multiplicities, while the full AM uses $x^\mathrm{ex}_\subj$ that includes the multiplicities. 
}
\label{tab:AUC}
\end{table}

\begin{table}
\begin{center}
\begin{tabular}{lcccccccc}
\toprule
\multirow{2}{*}{Model}
& 
\multicolumn{2}{c}{$\PT$ vs.~$\PQ$} & 
\multicolumn{2}{c}{$\HT$ vs.~$\HQ$} & 
\multicolumn{2}{c}{$\PQ$ vs.~$\HQ$}
& 
\multicolumn{2}{c}{$\PT$ vs.~$\HT$} 
\\
\cmidrule(lr){2-3}\cmidrule(lr){4-5}\cmidrule(lr){6-7}\cmidrule(lr){8-9}
&
$R_{50\%}$ & $R_{30\%}$ &
$R_{50\%}$ & $R_{30\%}$ &
$R_{50\%}$ & $R_{30\%}$ &
$R_{50\%}$ & $R_{30\%}$ \\
\midrule
CNN \cite{Chakraborty:2020yfc} & 
79.0 & 318 & 
55.9 & 219 &
2.33 & 4.16 &
2.57 & 4.83
\\
\cmidrule[0.2pt](lr){1-9}
\small{\emph{IRC-safe AM-PIPs}}
\\
A) $x_{\rm kin}$, $x_{S_2}$ &
71.1 & 276 &
49.1 & 186 &
2.51 & 4.63 &
2.48 & 4.49
\\
B) $x_{\rm kin}$, $x_\subj$ &
75.3 & 290 &
52.4 & 203 &
2.53 & 4.75 &
2.58 & 4.74
\\
C) $x_{\rm kin}$, $x_{S_2}$, $x_\subj$ &
80.7 & 316 &
56.0 & 217 &
2.59 & 4.87 &
2.64 & 4.85
\\
\cmidrule[0.2pt](lr){1-9}
\small{\emph{IRC-unsafe AM-PIPs}} &
\\
D) $x_{\rm kin}$, $x_{\rm MF}$, $x_{\rm count}$ &
57.8 & 215 &
40.8 & 152 &
2.61 & 4.96 &
2.65 & 4.96
\\
E) D) + $x_{S_2}$ &
81.2 & 321 &
56.5 & 223 &
2.74 & 5.31 &
2.76 & 5.19
\\
F) D) + $x_{S_2}$, $x_{\subj}$ &
84.8 & 341 &
58.9 & 234 &
2.76 & 5.36 &
2.78 & 5.21
\\
Full AM & 
85.7 & 343 & 
61.3 & \bf 244 & 
\bf 2.78 & \bf 5.43 &
2.82 & 5.36 
\\
\cmidrule[0.2pt](lr){1-9}
ParT \cite{Qu:2022mxj} & 
\bf 90.5 & \bf 372 & 
\bf 62.6 &  242 &
2.77 & 5.38 &
\bf 3.07 & \bf 5.94 
\\
\bottomrule
\end{tabular}
\end{center}
\caption{
Rejection rates $R_{X\%}$ for CNN, ParT, AM, and various configurations of AM-PIPs. 
The specific inputs activated in each AM-PIP are listed in the left column. 
Note that $x_\subj$ does not include subjet constituent multiplicities, while the full AM uses $x^\mathrm{ex}_\subj$ that includes the multiplicities. 
}
\label{tab:rejection}
\end{table}

We present the classification performances of AMs and ParT in terms of AUC\footnote{AUC stands for Area Under the receiver operator characteristic Curve.} in \tabref{tab:AUC}, and two rejection rates $R_{50\%}$ and $R_{30\%}$ in \tabref{tab:rejection}.
The rejection rate $R_{X\%}$ defined as follows:
\begin{equation}
R_{X\%} = 1/\fpr\ \mathrm{at\  TPR}=X\% ,\label{defRX}
\end{equation}
where TPR is the true positive rate, and FPR is the false positive rate. 
In this comparison, the first class is assumed to be the positive class, and the second class is considered the negative class. 
We report the performance metrics of AM and ParT that achieved the best AUC, selected from six AM and ten ParT trainings, each of them trained with different random seeds, respectively.

In \tabref{tab:AUC}, we see that AM performs almost as well as ParT in our classification tasks, except for the ($\PT$~vs.~$\HT$) classification.
We will discuss the significance of the differences in ($\PT$~vs.~$\HT$) later.
The AUCs for top jet tagging are similar, and AM's AUC is even a bit higher in ($\PQ$~vs.~$\HQ$), even though it only uses a few HLFs.
However, when AUCs are close, the corresponding ROC curves may intersect at some point. 
AUC alone may not clearly determine the performance hierarchy in these cases.

The rejection rates in \tabref{tab:rejection} are more precise measures for the performance comparison.
The rejection rates of ($\PT$~vs.~$\PQ$), ($\PQ$~vs.~$\HQ$), and ($\PT$~vs.~$\HT$) show the same trends seen in the AUCs. 
But the HW top tagging problem ($\HT$~vs.~$\HQ$), whose AUC was higher for ParT, now shows a mixed behavior: $R_{50\%}$ is better for ParT, $R_{30\%}$ is better for AM.
In summary, AM and ParT show similar performances in ($\HT$~vs.~$\HQ$), ParT slightly outperforms AM in ($\PT$~vs.~$\HT$) and ($\PT$~vs.~$\PQ$), but AM does slightly better in ($\PQ$~vs.~$\HQ$).
These experiments indicate that while AM and ParT are comparable in performance, their exact hierarchy is unclear. 
In \secref{sec:ex.margin}, we will discuss the significance of the differences using bootstrap methods, but the statistical significance after counting statistical uncertainty and training stochasticity does not exceed 3$\sigma$.

We tested several modifications of the AM, especially for closing the small gap between AM and ParT in the PY top tagging problem ($\PT$ vs.~$\PQ$), although HW top tagging and VIN top tagging (see \appref{subapp:vincia}) do not have such a performance gap. 
The modification includes various adjustments such as replacing the recursive network model for $x^\mathrm{ex}_\subj$ to a fully connected network, adding MFs at an energy threshold of 16~GeV, increasing the hidden layer numbers from two to three, and concatenating $x_\kin$ in each layer for more explicit jet kinematics conditioning.
However, these changes were insufficient to close the AUC gap.
The gap exclusive to PY suggests that introducing PY-specific HLFs appearing above the hadronic calorimeter resolution could potentially close it.
However, delving into HLFs specific to a particular Monte-Carlo simulation is beyond the scope of the paper, so we do not further attempt to fill the gap within this paper.

\subsubsection{Importance of each Sub-module of Analysis Model}

In addition to the direct comparison between AM and ParT, we further consider AM with partial inputs (AM-PIPs) for a hierarchical comparison of the inputs. 
Our AM incorporates selected sets of HLFs described in \secref{subsec:AMandHL}.
These HLFs, which include jet kinematics $x_{\kin}$, two-point energy correlations $x_{S_2}$, $p_T$-conditioned constituent multiplicity $x_\mathrm{count}$, Minkowski functionals for jet morphology $x_\mathrm{MF}$, subjet constituent multiplicities and kinematics $x^\mathrm{ex}_\mathrm{subj}$, are designed to capture various physics features of top jets illustrated in \figref{fig:top_anatomy}.
To evaluate the impact of each HLF set on the classification performance, we systematically activate or deactivate specific sets of inputs and their corresponding analysis modules. 
This method of toggling inputs allows us a hierarchical comparison, assessing the relative efficacy of these subsets within the complete AM framework.

Besides AMs and ParT, we consider a convolutional neural network (CNN) introduced in our previous works \cite{Chakraborty:2020yfc, Lim:2020igi}. 
The CNN demonstrated performance similar to that of AM-PIPs with inputs ($x_\kin$, $x_{S_2}$, $x_\mathrm{MF}$).
We will use its classification performances as a baseline for the following comparisons.

\tabref{tab:AUC} and \tabref {tab:rejection} also shows the results of AM-PIPs.
The tested input configurations are: A) ($x_{\rm kin}$,  $x_{S_2}$), B) ($x_{\rm kin}$, $x_\subj$), C) ($x_{\rm kin}$, $x_{S_2}, x_\subj$), and D) ($x_{\rm kin}$, $x_{\rm MF}$, $x_{\rm count}$), E) ($x_{\rm kin}$, $x_{S_2}$, $x_{\rm MF}$, $x_{\rm count}$), and F) ($x_{\rm kin}$, $x_{S_2}$,  $x_\subj$, $x_{\MF}$, $x_{\rm count}$). 
These configurations allow us to assess the impact of different HLF combinations on the classification performance.
Full AM outperforms all AM-PIPs, indicating that all the HLF sets contain independent information that AM can utilize effectively. 
Furthermore, the consistent improvement observed with the additional HLFs indicates that the statistical uncertainties are less significant in the performance comparison among AM-PIPs.

The first three AM-PIPs, configuration A), B), and C), omit the counting variables, making them less sensitive to the IRC-unsafe features. 
Each sub-module alone demonstrates effectiveness in the classifications, as shown in the A) and B) results.
But configuration C), combining the two-point energy correlations and subjet kinematics information, performs better, and its performance marginally surpasses that of CNN.
Note that CNN is essentially an extension of a local pointwise feature aggregator \cite{lin2013network}, and its design does not prioritize features from a certain location during aggregation.
In contrast, subjets have hierarchical structures.
Explicitly providing the hierarchical subjet information allows the AM-PIP to outperform CNN since we have provided additional information.

The latter three AM-PIPs, configuration D), E), and F), discuss IRC-unsafe features.
We first observe that the performance drops significantly in the top tagging problems ($\PQ$~vs.~$\HQ$) and ($\PT$~vs.~$\HT$) if we restrict AM to access only to constituent multiplicities and topological and morphological features as in the configuration D). 
The performance is even worse than CNN.
Note that MFs are geometric measures, so they are specialized in measuring sizes rather than measuring relative positions, even though scale-dependent MF analysis helps analyze the relative structures at some level.
Identifying the local clusters and their relative structure is very important in top tagging, so CNN performs better than the configuration D).
These observations show that IRC safe features for analyzing the structure of jets, such as $x_{S_2}$ and $x_\subj$, are important in top tagging problems using samples from the same generators.

But in the case of comparing simulated jets from different Monte Carlo simulations, ($\PT$~vs.~$\PQ$) and ($\HT$~vs.~$\HQ$), the setup D) starts to outperform the IRC-safe AM-PIPs.
The difference between simulated jets is more prominent in IRC-unsafe features because they have significant simulation model dependence, so D) can distinguish jets from different simulations more clearly than the others. 
Considering this, orthogonal information improves the performance.
The setup D) is even performing better than CNN in this case, contrary to what we observed in the top tagging problems.
The overall geometry of particle distribution is more important than hard substructures for distinguishing Monte Carlo simulations.
Although CNN structure can cover MFs \cite{Lim:2020igi} in the large depth limit, using MFs explicitly could help gain performances compared to CNN with moderate depth.

The last comparison among E), F), full AM, and ParT justifies the importance of subjet kinematics and their color structures.
The baseline of this comparison is E), which is the AM-PIP without subjet information.
Including subjet kinematic information $x_{\subj}$ to E) slightly improves performance, as seen in F).
The reason is that two-point energy correlations, counting observables, and Minkowski functionals treat all the constituents equally during the feature aggregation while subjet information is not.
However, the performance of F) still does not closely match that of ParT.
Further considering subjet constituent multiplicities makes F) to the full AM, whose performance is comparable to ParT as discussed in the previous sections.
This shows the colors of subjets, i.e., the origin of the subjets ($q$~or~$g$), plays a certain role in the classifications.  

\section{Statistical Significance of the Difference between Analysis Model and Particle Transformer}
\label{sec:ex.margin}

\begin{table}
\begin{center}
\begin{tabular}{lccc}
\toprule
\multirow{2}{*}{Model}
& 
\multicolumn{3}{c}{$\PT$ vs.~$\PQ$} 
\\
\cmidrule(lr){2-4}
& 
AUC & $R_{50\%}$ & $R_{30\%}$
\\
\midrule
ParT-best (as I) &
0.9425$\pm$0.0005 &
85$\pm$2($\pm1$)&
327$\pm$12($\pm9$) 
\\
ParT-2nd (as II) &
0.942\phantom{0}$\pm$0.001\phantom{0} &
83$\pm$3($\pm1$) &
317$\pm$14($\pm9$)  
\\
AM (as II) &
0.9396$\pm$0.0006 &
77$\pm$2($\pm1$) &
296$\pm$12($\pm8$) 
\\
\cmidrule[1pt](){1-4}
\multirow{2}{*}{Model}
& 
\multicolumn{3}{c}{$\HT$ vs.~$\HQ$} 
\\
\cmidrule(lr){2-4}
& 
AUC & $R_{50\%}$ & $R_{30\%}$ 
\\
\midrule
ParT-best (as I) &
0.9266$\pm$0.0009 &
59\phantom{.0}$\pm$2\phantom{.0}($\pm0.7$) &
222$\pm$10($\pm5$)
\\
ParT-2nd (as II) &
0.9265$\pm$0.0008 &
58\phantom{.0}$\pm$2\phantom{.0}($\pm0.7$) &
225$\pm$\phantom{0}8($\pm5$)
\\
AM (as II) &
0.9240$\pm$0.0005 &
55.2$\pm$0.7($\pm0.6$)&
218$\pm$\phantom{0}6($\pm5$)
\\
\cmidrule[1pt](){1-4}
\multirow{2}{*}{Model}
& 
\multicolumn{3}{c}{$\PQ$ vs.~$\HQ$}
\\
\cmidrule(lr){2-4}
& 
AUC & $R_{50\%}$ & $R_{30\%}$ 
\\
\midrule
ParT-best (as I) &
0.592\phantom{0}$\pm$0.004\phantom{0} &
2.70\phantom{0}$\pm$0.04\phantom{0}($\pm0.006$) &
5.2\phantom{0}$\pm$0.1\phantom{0}($\pm0.02$) 
\\
ParT-2nd (as II) &
0.591\phantom{0}$\pm$0.004\phantom{0} &
2.70\phantom{0}$\pm$0.04\phantom{0}($\pm0.006$) &
5.14$\pm$0.08($\pm0.02$)
\\
AM (as II) &
0.5949$\pm$0.0006 &
2.732$\pm$0.005($\pm0.006$) &
5.27$\pm$0.02($\pm0.02$) 
\\
\cmidrule[1pt](){1-4}
\multirow{2}{*}{Model}
& 
\multicolumn{3}{c}{$\PT$ vs.~$\HT$} 
\\
\cmidrule(lr){2-4}
& 
AUC & $R_{50\%}$ & $R_{30\%}$ 
\\
\midrule
ParT-best (as I) &
0.614$\pm$0.006 &
2.92$\pm$0.06($\pm0.01$) &
5.6\phantom{0}$\pm$0.1\phantom{0}($\pm0.02$)
\\
ParT-2nd (as II) &
0.613$\pm$0.007 &
2.91$\pm$0.08($\pm0.01$) &
5.6\phantom{0}$\pm$0.2\phantom{0}($\pm0.02$)
\\
AM (as II) &
0.598$\pm$0.001 &
2.75$\pm$0.01($\pm0.01$) &
5.17$\pm$0.02($\pm0.02$)
\\
\bottomrule
\end{tabular}
\end{center}
\caption{
AUCs and rejection rates $R_{X\%}$ of the AM and ParT, both trained on bootstrap datasets. 
We show the average and standard deviations of these classification performance metrics.
For ParT, the classifier is trained twice; the ParT with a higher average AUC is labeled as ParT-best, and the other as ParT-2nd. 
The numbers in brackets represent the statistical uncertainties associated with each metric. 
Note that the bootstrap estimated uncertainties already include this.
}
\label{Table:bootParT}
\end{table}

In \secref{sec:compare}, we observed that 1) AM and ParT performance are comparable, and 2) AM-PIP improves as more inputs are included. 
In this section, we estimate the statistical uncertainties of the performance measures to show the significance of our conclusion.

\subsection{Estimating Uncertainty of Performance Metrics by Bootstrapping}
\label{sec:uncertainty}

We estimate uncertainties associated with network training by using the bootstrap method.
This method involves the generation of bootstrap datasets from the original dataset by sampling with replacement.
In this resampling procedure, we treat the empirical distribution of the training dataset as an approximate true distribution.
The bootstrap datasets effectively act as representative samples of this approximate true distribution.
By training networks again on each of these datasets, this bootstrap method allows us to estimate the variance of training results.
For this study, we use ten bootstrap datasets.
Note that our AM is computationally cheaper than ParT, so this bootstrap analysis can be done much faster.
The computational resources utilized by AM and ParT are explained in \appref{sec:scomp}.

In \tabref{Table:bootParT}, we show the average and standard deviation of AUC, $R_{50\%}$, and $R_{30\%}$ of AMs and ParTs trained on the bootstrap dataset.
Here, we use the original test datasets without bootstrapping to evaluate those performance measures. 
Note that the performance measures are worse than those in \tabref{tab:AUC} and \tabref{tab:rejection}.
The difference is due to the bootstrap dataset being sampled from the original dataset.
The resampling does not capture all the training samples, which makes the classifiers slightly suboptimal.
Nevertheless, the variance is still a consistent estimator of the uncertainty.
For ParT, we further validate this method by repeating the procedure with another set of bootstrap datasets. 
We denote the bootstrap result with a higher average AUC as the ParT-best and the other as ParT-2nd.
The ParT-best and ParT-2nd results are compatible with the uncertainties.

One notable observation is that the bootstrap-estimated uncertainties for ParT are significantly larger than those for AM.  
Specifically, in the HW top tagging problem, the uncertainties of ParT are twice as large as those of AM. 
The uncertainty difference is amplified by factor 10 in the generator classifications ($\PT$~vs.~$\HT$) and ($\PQ$~vs.~$\HQ$).
This phenomenon is due to the bias-variance tradeoff inherent in statistical modeling.
More expressive models, like ParT, can model classifiers more accurately, but at the cost of increased variance in the results.
Consequently, ParT requires a larger sample size to achieve the same level of statistical precision as AM; for an analysis with a given sample size, the uncertainty of AM is always expected to be smaller than that of ParT.

This precision difference becomes more prominent if we subtract the statistical uncertainty from the bootstrap estimated uncertainty.
Here, we specifically focus on rejection rates, as their statistical uncertainty can be quickly estimated using analytic formulas.
A~naive estimation on $1\sigma$ statistical uncertainty of the false positive rate $R_{X\%}^{-1}$ is given by the Wald interval, $\sigma_{R_{X\%}^{-1}} = \sqrt{R_{X\%}^{-1}(1-R_{X\%}^{-1}) / N}$, where $N$ is the number of second class samples.
The uncertainty in the rejection rate, $\sigma_{R_{X\%}}$, is then derive as $\sigma_{R_{X\%}} = {\sigma_{R_{X\%}^{-1}}} / {R_{X\%}^2}$.
The statistical uncertainty estimates for the rejection rates are listed in \tabref{Table:bootParT}, denoted within brackets.

We can see that the uncertainty is dominated by the statistical uncertainty in the case of AMs, whereas for ParT, this is not the case.
This additional variance observed in ParT is from randomness of training, such as network initializations and batch splitting.
In the case of AM, the training procedure sufficiently suppresses this initial condition and other randomness dependence in training so that the bootstrap estimated uncertainty and statistical uncertainty are approximately the same.
In contrast, ParT training is less successful in reducing this variance to levels below the statistical uncertainties, resulting in higher residual variance in the bootstrap estimated uncertainties.
This stability in the training process is one significant advantage of AMs, especially when their performance is comparable to that of state-of-the-art models like ParT.

\begin{table}[t!]
    \centering
    \begin{tabular}{ccccc}
\toprule
        metric& $\PT$~vs.~$\PQ$& $\HT$ vs.~$\HQ$ & $\PQ$ vs.~$\HQ$ & $\PT$ vs.~$\HT$  \\
\midrule
AUC & 2.76 &  2.58& 0.84 & 2.36\\
$R_{50\%}$ &  2.19 & 1.56 & 0.79  & 2.29 \\
$R_{30\%}$ &  1.64  & 0.51 & 1.08  &2.84\\
\bottomrule
    \end{tabular}
    \caption{Estimated significance of the difference between classifier performance metrics between AM and ParT, based on the training uncertainties measured by the bootstrap method. }
    \label{tab:sig}
\end{table}

For a more quantitative comparison of the difference in performance measures between AM and ParT, we define the significance of their performance measure difference as follows:
\begin{equation}
\sigma_X=\frac{\vert X_\ParT-X_\AM\vert}{\sqrt{\sigma_{X,\ParT}^2+\sigma_{X,\AM}^2}}, \quad X \in \{\mathrm{AUC}, R_{50\%}, R_{30\%} \}.
\end{equation}
For $X_{\mathrm{ParT}}$ and $\sigma_{X,\mathrm{ParT}}$, we use the average of ParT-best and Part-2nd results.
The significances $\sigma_{X}$ are shown in \tabref{tab:sig}, all of which are less than $3\sigma$.
This indicates that the moderate performance gap between AM and ParT in PY top tagging ($\PT$~vs.~$\PQ$) and top jet simulation comparison ($\PT$ vs.~$\HT$), shown in \tabref{tab:AUC} and \tabref{tab:rejection}, becomes less significant after accounting both statistical uncertainty and training stochasticity.

\begin{figure}[t!]
    \centering
    \includegraphics[width=0.6\textwidth]{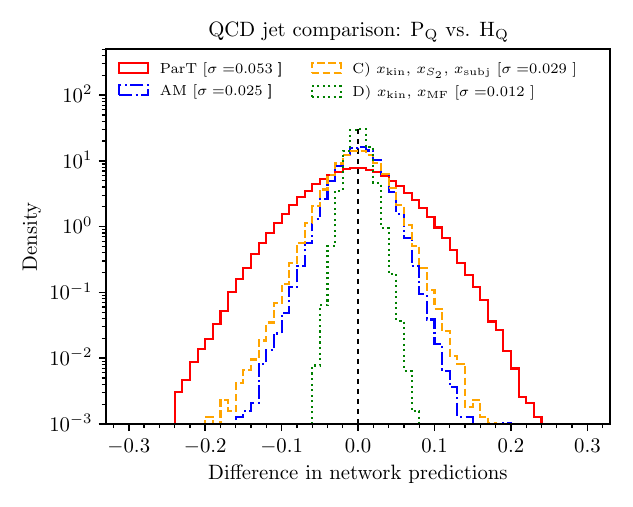}
    \caption{
    Distribution of the difference between the classifier outputs of the best model and the second best model for $\PQ$~vs.~$\HQ$ classification.
    The red solid line is for ParT, and the blue dot-dashed line is for AM.
    The distribution for ParT is significantly broader than that of AM.
    Additionally, the green dotted line is for the AM-PIP with only the morphology analysis module active, and the orange dashed line is for the AM-PIP with relation network and subjet analysis modules active.
    These two extra lines are for demonstrating output variations in simpler models compared to the full AM.
    }
    \label{fig:diff}
\end{figure}

Finally, to test the consistency of AM and ParT predictions under the randomness of the training, we revisit the classifiers for ($\PQ$~vs.~$\HQ$) trained with different random seeds, as discussed in \secref{sec:comp:am_part}.
\figref{fig:diff} illustrates the differences in classifier outputs between the networks with the best and second-best AUCs.
A classifier less influenced by training randomness will yield more similar network outputs.
The difference histogram will peak narrowly around zero in such a case.
In this comparison, the peak for AM (red dot-dashed line) is narrower than that for ParT (red solid line), indicating more consistent predictions.
Specifically, the standard deviation of the histogram is 0.03 for AM, and 0.06 for ParT. 
A simpler model, such as AM-PIP involving $(x_\kin, x_\MF, x_\rcount)$, shows even smaller variation of 0.012, though its performance is suboptimal.

\subsection{Comparing the Information Content of Classifiers using Odds Ratio}

In this section, we directly compare AM and ParT using an odds ratio, which is useful in comparing the information content of two classifiers.
Let $s^{\I}(x)$ be the classifier output for given input $x$, i.e., the posterior being a signal as modeled by classifier ${\I}$.
The odds ratio $\hat{w}^{\I/\II}$ is defined as the ratio of odds estimated by models $\I$ and $\II$, expressed as follows:
\begin{eqnarray}
    \mathrm{OR}^{\I/\II}
    (x)
    &=&
    \frac
    {s^{\I}(x)}
    {1-s^{\I}(x)}
    \cdot
    \frac
    {1-s^{\II}(x)}
    {s^{\II}(x)}.\label{margew}
\label{eq:oddI}
\end{eqnarray}
An odds ratio greater than 1 indicates that the classifier $\I$ is more confident in its signal prediction by the factor of the odds ratio than the classifier $\II$.
If both models are identical, the odds ratio will be 1.\footnote{Note that this reference point 1 may change if you compare classifiers trained using different class priors. Here, we only consider classifiers trained using class prior probability 0.5. }
Therefore, deviations from 1 in the odds ratio indicate differences in the predictions of the two classifiers for a given sample.

Note that when there is a clear hierarchy between two models, the odds ratio can also be interpreted as a likelihood ratio between two classes, using only the features exclusive to the more expressive model.
If the classifier $\I$ is more expressive than $\II$, the prediction $s^{\II}(x)$ in the denominator acts as a normalizer countering information in $s^{\I}(x)$ covered by the model $\II$, similar to conditional probability.
With this odds ratio interpretation, we can construct a posterior probability that works as a classifier utilizing only the features exclusive to $\I$:
\begin{eqnarray}
    s^{\I/\II}(x)
    =
    \frac
    {\mathrm{OR}^{\I/\II}(x)}
    {1+\mathrm{OR}^{\I/\II}(x)}.
\label{eq:oddII}
\end{eqnarray}
A more detailed derivation of this interpretation is described in \appref{app:proof_s_marg}.

Since the odds ratio has a classifier interpretation, we quantify the difference between AM and ParT in terms of performance metrics (AUC and rejection rates) of ROC curves using the probability $s^{\I/\II}$ (equivalently the odds ratio $\mathrm{OR}^{\I/\II}$) as a threshold variable.
If two classifiers are equivalent, the AUC will be that of a random guess, 0.5, and the rejection rate $R_{X\%}$ will be $1/X\%$.

To estimate uncertainties together, we take the average prediction of the ParT-best models from the bootstrap analysis as ${s}^\I$.
We then evaluate the odds ratio against AM and ParT-2nd, each trained on bootstrap datasets.
Specifically, for each of the $i$-th bootstrap dataset, $s^{\II}$ is either the prediction of AM, $s^\AM_i$, or ParT-2nd, $s^{\textrm{ParT-2nd}}_i$.
We evaluate the mean and variance of AUC and rejection rates ($R_{50\%}$ and $R_{30\%}$) across the bootstrap datasets to assess the difference between the two models.

The above uncertainty estimation procedure primarily focuses on model $\II$ variations and does not explicitly account for the uncertainties of model $\I$.
To estimate the uncertainties in ParT-best, we can take ParT-2nd as $s^{\II}$.
The uncertainties associated with ParT-2nd can be interpreted as an additional uncertainty due to the variation within ParT-best as model $\I$ because ParT-best and ParT-2nd are essentially the same models trained under different random states.

\begin{table}[t!]
    \centering
    \begin{tabular}{cccccc}
\toprule
        $\II$& metric& $\PT$ vs.~$\PQ$& $\HT$ vs.~$\HQ$ & $\PQ$ vs.~$\HQ$ & $\PT$ vs.~$\HT$  \\
\midrule
ParT-2nd 
& AUC & 0.52$\pm$0.05 &  0.53$\pm$0.04 & 0.50$\pm$0.02 & 0.50$\pm$0.02\\
&$R_{50\%}$ &  2.2$\pm$0.4 & 2.2$\pm$0.3 & 2.03$\pm$0.10   & 2.0$\pm$0.1 \\
&$R_{30\%}$ & 4.2$\pm$0.9 & 4.3$\pm$0.9 & 3.4$\pm$0.2  & 3.3$\pm$0.3\\
\midrule
AM 
& AUC & 0.48$\pm$0.02 & 0.47$\pm$0.01 & 0.507$\pm$0.003 & 0.546$\pm$0.003 \\
&$R_{50\%}$ & 1.87$\pm$0.09 & 1.83$\pm$0.05 & 2.04$\pm$0.02  & 2.29$\pm$0.02 \\
&$R_{30\%}$ & 3.2$\pm$0.2 & 3.2$\pm$0.2 & 3.41$\pm$0.04  & 3.91$\pm$0.05\\
\bottomrule
    \end{tabular}
\caption{ 
Classification performance metrics for the odds ratio $\mathrm{OR}^{\I/\II}$, where $\I$ is ParT-best and $\II$ is AM and ParT-2nd.
This table uses the results from bootstrap analysis, and we show the training uncertainties together.
An AUC value closer to 0.5 and rejection rates $R_{X\%}$ closer to $1/X\%$ indicates that the two classifiers have similar information contents.
}
\label{Table:marginbootParT}
\end{table}

In \tabref{Table:marginbootParT}, we show the odds ratio analysis results.
Ideally, the results of ParT-2nd should be those of random guess results ($\mathrm{AUC}=0.5$, $R_{50\%}=2$, and $R_{30\%}=3.33$). 
The rejection rates $R_{30\%}$ for the top tagging problems ($\PT$~vs.~$\PQ$) and ($\HT$~vs.~$\HQ$) are slightly deviate from 3.33. 
However, the uncertainty is also sufficiently large, so the results are consistent with the expectation.
Overall, all the performance metrics are compatible with those of a random guess, and the expectation holds for all the classification tasks.

The results of AM apparently deviated from those of a random guess, especially in ($\PT$~vs.~$\HT$).
The AUC for this classification is $0.546 \pm 0.003$. 
However, it is important to remember that the uncertainties of AM's performance metric are relatively small compared to those of ParT.
In this odds ratio analysis between ParT and AM, the uncertainties are dominated by those of ParT-best.
When comparing 0.046 to that in the ParT-2nd analysis, 0.02, the deviation is about twice the uncertainty.
Similarly, when considering the uncertainties of ParT-best, all the performance metrics for AM are compatible with random guess within approximately $2\sigma$.

\begin{figure*}[t!]
\centering
\includegraphics[width=0.495\textwidth]{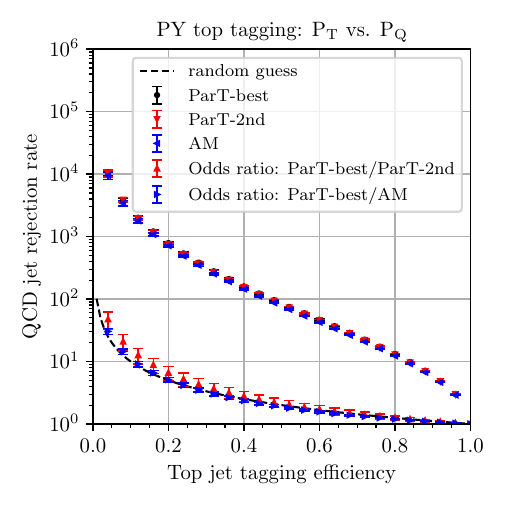}
\includegraphics[width=0.495\textwidth]{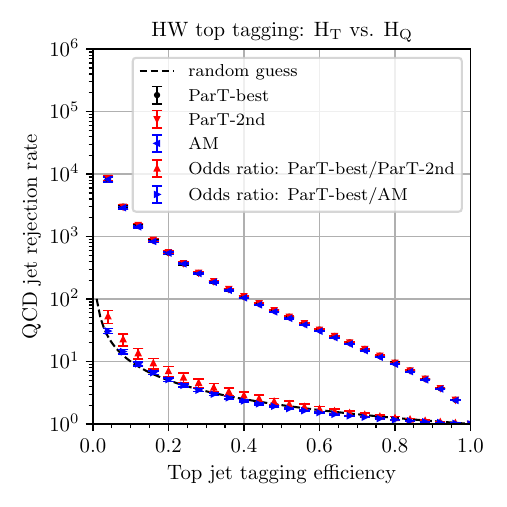}
\includegraphics[width=0.495\textwidth]{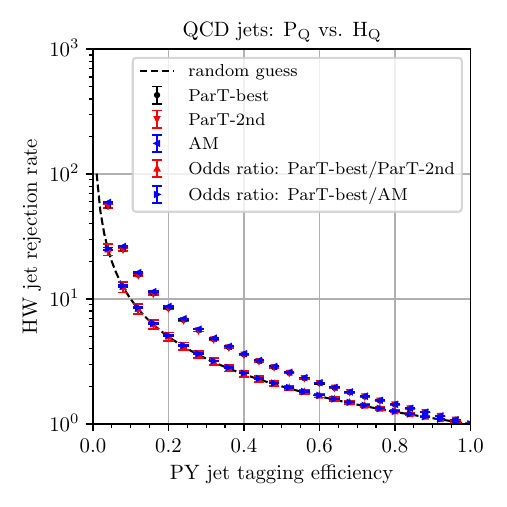}
\includegraphics[width=0.495\textwidth]{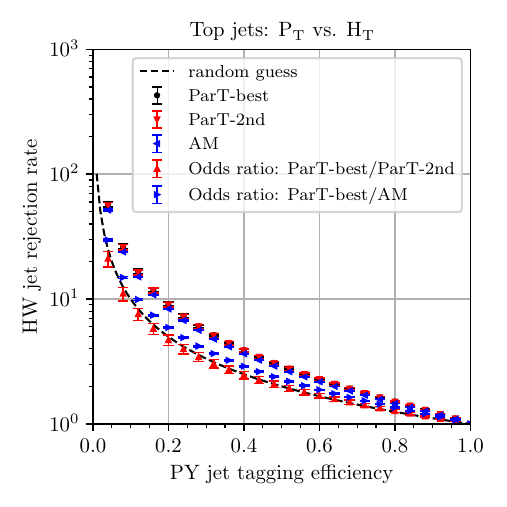}
\caption{
The average ROC curves and their standard deviations for the AM, ParT-best, and ParT-2nd trained on bootstrap datasets. 
Additionally, we show the ROC curves for the odds ratios $\mathrm{OR}^{\I/\II}$, where $\I$ is the ensemble average of ParT-best, and $\II$ is AM or ParT-2nd.  
The ROC curves shown here are for the following classification problems: 
(top~left)~$\PT$~vs.~$\PQ$, (top~right)~$\HT$~vs.~$\HQ$, (bottom~left)~$\PQ$~vs.~$\HQ$, and (bottom~right)~$\PT$~vs.~$\HT$. 
}
\label{Fig:marginbootParT}
\end{figure*}

Finally, \figref{Fig:marginbootParT} shows the average and standard deviations of the ROC curves for the following classifiers trained on the bootstrap datasets: ParT-best (black circle), ParT-2nd (red inverted triangle), and AM (blue left-facing triangle).
Additionally, we show the ROC curves from the odds ratio analysis \eqref{eq:oddI} and \eqref{eq:oddII} of ParT-best vs.~ParT-2nd (in red triangles) and ParT-best against AM (in blue right-facing triangles). 
The ROC curves of AM are compatible with that of ParT within $3\sigma$, consistent with other metrics discussed previously.
The ROC curves in the odds ratio analysis between ParT-best and AM look closer to the random guess, except for the classification ($\PT$~vs.~$\HT$).
But again, comparing the difference to the uncertainties of the odds ratio between ParT-best and Part-2nd, the gap becomes less significant.
In summary, our AM is aligned to ParT sufficiently well for the top jet tagging and distinguishing simulations while showing smaller training uncertainties.

\section{Discussion on Analysis Model and High-Level Features at Sub-Hadronic Calorimeter Angular Resolution}
\label{sec:ecal_scale}

Throughout the previous analyses, we used datasets pixelated or binned to the angular resolution of the hadronic calorimeter (HCAL), $\Delta R_H \simeq 0.1$.
The classification performance of our AM is consistent with one of the state-of-the-art models, Particle Transformer. 
A natural follow-up question is whether this consistent performance still holds at a finer resolution, such as the angular resolution of the electromagnetic calorimeter (ECAL), $\Delta R_E \simeq 0.025$.
This section addresses the expected AM performance at sub-HCAL angular resolution.

Before delving into the analysis of sub-HCAL angular resolution, we must pay attention to new features appearing at the smaller scales.
The first important difference is that ECAL towers and tracks are implicitly distinguishable from HCAL towers in the EFlow object analysis at sub-HCAL angular resolution.
If there is a constituent cluster with a size below $\Delta R_H$, that cluster is highly likely to consist of ECAL towers and tracks. 
Similarly, if a cluster is smaller than $\Delta R_E$, the cluster mostly consists of tracks.

This implicit accessibility of constituent type leads to an interesting consequence: particle identification (PID) information is also partially and implicitly accessible in sub-HCAL angular resolution analysis.
Comparing energy deposits in the HCAL, ECAL, and tracks partially reveals how much energy is carried by certain types of particles.
This is useful information when comparing jets with different PID compositions, and the use of the energy comparison has been studied in the context of jet flavor identification \cite{Park:2017rfb, Nakai:2020kuu}.
Also, explicitly considering PID information improves jet tagging performances, and it has been demonstrated using particle level analysis, i.e., when there is no implicit utilization of angular resolution \cite{Qu:2019gqs}.

Designing an analysis module and high-level features for capturing this constituent type information may further supercharge the tagging performance of AM.
However, constructing a new module analyzing implicit PID information is beyond the scope of the paper.
We will focus on tweaking the current AM setup to a finer resolution analysis and compare its performance to that of ParT.

For adjusting AM for higher resolution, we first have to be cautious about an inflated number of inputs. 
For example, the set of two-point energy correlations with trimming, $x_{\mathrm{trim}}$, described in \secref{subsec:pIII}, originally consisted of 60 inputs, i.e., (3 types of $S_{2,ab}$) $\times$ (20 bins). 
But at the ECAL angular resolution level, now we have to consider 80 bins per one $S_{2,ab}$, and it inflates the number of inputs by a factor of 4.
Similarly, the set of MFs, $x_{\mathrm{MF}}$, described in \secref{subsec:pIV}, consists of 150 inputs, i.e.,
(3 Minkowski functionals) $\times$ (10 dilation scales) $\times$ (5 $p_T$ thresholds).
Again, we have to consider four times more dilation scales to cover the same dilation scale range, so the total number of inputs will be inflated to 600.

The reason why we have to be cautious of inflated inputs is that it may make the training dataset more sparsely distributed on the input space. 
Jets in our dataset contain about 50 constituents, and the dimensionality of the input space of the AM for ECAL angular resolution analysis exceeds the dimension of the 4-momenta of jet constituents, $4 \times 50$.
In this case, less carefully tuned AM may require more training samples to compensate for the sparsity of the data.

Here, we present the results of AM tweaked for the ECAL angular resolution $\Delta R_E$. 
In this quick analysis, instead of using finer pixels and bins everywhere, we simply augment the previous AM for HCAL angular resolution analysis in order to avoid input dimension inflation, making the dataset sparse.
Here is the list of changes:
\begin{itemize}
    \item Pixelated jet constituent multiplicity variables in $x_{\rm count}$, $x_{\rm subj}$ are now evaluated using jet images with a pixel size $\Delta R_E \times \Delta R_E$.
    \item Two-point energy correlation analysis modules $\Phi^{\mathrm{trim}}$ and $\Phi^{\mathrm{trim}}$ additionally takes extra binned $S_{2,ab}$'s on a domain $[0, 0.25)$ and with bin size 0.025.
    \item Minkowski functional analysis module $\Phi^{\mathrm{MF}}$ additionally accounts Minkowski functionals with dilation scales in $[0.0125,0.125]$ and with step size 0.0125.
\end{itemize}
Namely, we count the sub-HCAL angular scale information up to the angular scale $R=0.25$, and we consider information above the ECAL scale only using inputs with HCAL angular resolution.
With this setup, each of the analysis modules is able to take care of differences in their high-level inputs in ECAL and HCAL angular resolution around and below HCAL angular resolution.
Also, 10 new inputs per each of $S_{2,ab}$ and MFs are considered so that the input dimensions are not inflated too much.

The classification performances of AM and ParT at this ECAL angular scale are shown in \tabref{tab:0025AUC} and \tabref{tab:0025rejection}. 
Here, the average and uncertainties of the performance metrics are estimated using a bootstrap method. 
The ParT model used here is the same ParT model described in \secref{subsec:ParT}.
The difference in the classification metrics between AM and ParT lies between $2\sigma$ to $5\sigma$ when we only count variation of classifier under change of random seeds.
Again, the error in this comparison is mostly dominated by uncertainties of ParT training, and $\PT$~vs.~$\HT$ shows the largest difference in metric about $5\sigma$.
We will leave accounting for the additional high-level features for the smaller scales, further fine-tuning AM setup and binning of the inputs,  and estimating the full uncertainty precisely for future studies.

\begin{table}
\begin{center}
\begin{tabular}{rcccc}
\toprule
\multirow{2}{*}{Model} & 
$\PT$ vs.~$\PQ$ & 
$\HT$ vs.~$\HQ$ & 
$\PQ$ vs.~$\HQ$ & 
$\PT$ vs.~$\HT$ 
\\
\cmidrule(lr){2-2}\cmidrule(lr){3-3}\cmidrule(lr){4-4}\cmidrule(lr){5-5}
& AUC & AUC & AUC & AUC
\\
\midrule
\cmidrule[0.2pt](lr){1-5}
Full AM&
0.9467$\pm$0.0002&
0.9313$\pm$0.0001 &
0.608$\pm$0.004&
0.618$\pm$0.001
\phantom{0}
\\
\cmidrule[0.2pt](lr){1-5}
ParT 
& 
0.9498$\pm$0.0011 &
0.9348$\pm$0.0007&
0.620$\pm$0.005 &
0.649$\pm$0.006
\phantom{0}
\\
\bottomrule
\end{tabular}
\end{center}
\caption{ The average and standard deviations of AUC, considering ten networks trained using different random number seeds and trained on bootstrapped samples. 
Here, we use the same ParT model described in \secref{subsec:ParT}.
}
\label{tab:0025AUC}
\end{table}

\begin{table}
\begin{center}
\begin{tabular}{rcccc}
\toprule
\multirow{2}{*}{Model}
& 
\multicolumn{2}{c}{$\PT$ vs.~$\PQ$} & 
\multicolumn{2}{c}{$\HT$ vs.~$\HQ$} 
\\
\cmidrule(lr){2-3}\cmidrule(lr){4-5}
&
$R_{50\%}$ & $R_{30\%}$ &
$R_{50\%}$ & $R_{30\%}$ \\
\midrule
Full AM & 
\phantom{0}96$\pm$1 & 392$\pm$\phantom{0}4 &
67.5$\pm$0.4 & 281$\pm$\phantom{0}6
\\
ParT & 
106$\pm$4 & 430$\pm$26 & 
73.9$\pm$2.0 & 302$\pm$13 
\\
\midrule
\multirow{2}{*}{Model}
& 
\multicolumn{2}{c}{$\PQ$ vs.~$\HQ$} & 
\multicolumn{2}{c}{$\PT$ vs.~$\HT$} 
\\
\cmidrule(lr){2-3}\cmidrule(lr){4-5}
&
$R_{50\%}$ & $R_{30\%}$ &
$R_{50\%}$ & $R_{30\%}$ \\
\midrule
Full AM & 
2.87$\pm$0.01& 5.61$\pm$0.03&
2.97$\pm$0.02 & 5.74$\pm$0.03
\\
ParT & 
3.01$\pm$0.06 & 5.93$\pm$0.15 &
3.38$\pm$0.08 & 6.77$\pm$0.20 
\\
\bottomrule
\end{tabular}
\end{center}
\caption{
The average and standard deviations of the rejection rates $R_{50\%}$ and $R_{30\%}$, considering ten networks trained using different random number seeds and trained on bootstrapped samples.
Here, we use the same ParT model described in \secref{subsec:ParT}.
}
\label{tab:0025rejection}
\end{table}

\section{Summary and Conclusion}
\label{sec:conclusion}

This paper introduces an analysis model (AM) employing the important high-level features (HLFs) in top jet tagging.
The AM simplifies the current state-of-the-art taggers analyzing low-level features (LLFs) while maintaining classification performance.
To demonstrate the effectiveness of our model, we revisited the problem of top jet tagging using various Monte Carlo simulated samples at the angular resolution of a hadronic calorimeter scale. 
We compared the classification performance of AM with a classifier based on the Particle Transformer (ParT).
In our previous works, we have developed AMs using the IRC-safe relation network \cite{Lim:2018toa, Chakraborty:2019imr, Chakraborty:2020yfc} and mathematical morphology for jet physics \cite{Chakraborty:2020yfc, Lim:2020igi}.  
This model's performance matched the performance of a convolutional neural network, but it fell short of matching ParT's performance.
Nevertheless, once we fully considered the newly introduced features -- {$p_T$-conditioned} jet constituent multiplicity distribution and subjet constituent multiplicities for capturing color charges of subjets -- the performance of AM is aligned with that of ParT within a range of $2\sim3\sigma$ of training uncertainties.

The compatibility between our AM and ParT varies depending on the specific classification problem.
For instance, in the top tagging problem using \texttt{Herwig}-generated samples, AM demonstrates comparable performance to that of ParT. 
Similarly, AM matches ParT's performance in distinguishing \texttt{Pythia}-generated QCD jets and \texttt{Herwig}-generated QCD jets.
In the top tagging problem with \texttt{Pythia}-generated samples and distinguishing \texttt{Pythia}-generated top jets and \texttt{Herwig}-generated top jets, we observed discrepancies at the level of $2\sim3\sigma$ of training uncertainty.
The results suggest that \texttt{Pythia}-simulated jets may have unique features that are not covered by AM, and those are absent in \texttt{Herwig}-simulated jets.
Identifying the HLFs exclusive to \texttt{Pythia} to fill this minor gap would be interesting, particularly in understanding the differences between simulations and calibrating those with experimental data; we plan to explore this in future studies.

One key advantage of employing simpler classifiers is the reduction in training uncertainty, a direct consequence of the bias-variance tradeoff.
We explicitly demonstrated this behavior in AM through bootstrap methods applied to our jet tagging problems, and our AM demonstrated much smaller training uncertainty than ParT.
Importantly, the smaller uncertainty is useful when classifiers are utilized for estimating statistical quantities, such as a likelihood ratio for reweighting simulated datasets \cite{nextpaper} and decorrelating certain features by planing \cite{Chang:2017kvc, deOliveira:2015xxd}.  
We expect that our AM and this kind of HLF-analyzing classifier-based approach will be valuable in narrowing down uncertainties of reweighting-based calibration for simulated events \cite{Andreassen:2019nnm, Gambhir:2022dut} and synthetic samples from machine-learned generative models \cite{Diefenbacher:2020rna,Butter:2021csz,Das:2023ktd}.

\acknowledgments

We thank David Shih and Edward Ramirez for the useful comments and discussions. This work is supported by Grant-in-Aid for Transformative Research Area (A) 22H05113 and Grant-in-Aid for Scientific Research(C) JSPS KAKENHI Grant Number 22K03629. The work of SHL was also supported by the US Department of Energy under grant DE-SC0010008. The authors acknowledge the Office of Advanced Research Computing (OARC) at Rutgers, The State University of New Jersey for providing access to the Amarel cluster and associated research computing resources that have contributed to the results reported here (URL:~\url{https://oarc.rutgers.edu}). This paper is revised using large language models,  \texttt{ChatGPT4} and \texttt{Claude 3 Opus}. The authors guarantee that the AI tools are used only to improve writing quality, not to generate the idea and paper itself.

\appendix

\section{Classification Results Involving Vincia Simulations}\label{subapp:vincia}

In this appendix, we present the classification results using samples generated \texttt{Vincia} (VIN). 
We remind that VIN-simulated top jet datasets and QCD jet datasets are denoted by $\VT$~and~$\VQ$, respectively.
Again, we focus on the following two types of classification problems: 
The first problem is top jet tagging trained on VIN-generated samples ($\VT$~vs.~$\VQ$).
The second problem is comparing VIN-simulated jets with jets from other simulations, specifically comparing PY and VIN (($\PQ$~vs.~$\VQ$) and ($\PT$ vs.~$\VT$)), and comparing VIN and HW (($\VQ$~vs.~$\HQ$) and ($\VT$ vs.~$\HT$)).
We do not perform bootstrap analysis in this appendix; however, we expect that the training uncertainties will be similar to those of the corresponding problem in the main text, given that the same AM and ParT are used.

The classification performance metrics are shown in \tabref{tab:vinauc}, 
\tabref{tab:vinrejectionI}, and \tabref{tab:vinrejectionII}. 
We show the naive statistical uncertainty estimations in the brackets.
The difference in the classification performances between AM and ParT is negligible, mostly falling within the total training uncertainty.
For the top jet simulation comparisons, some deviation is observed considering the statistical uncertainty.
However, the total training uncertainty is about 5 to 10 times larger considering \tabref{Table:bootParT}, and hence, the difference is not significant.

One notable observation is that the AUCs of top taggers trained on VIN-generated samples are larger than those of top taggers trained on PY-generated samples or HW-generated samples, shown in \tabref{tab:AUC}.
This suggests that the systematical uncertainties from event generators remain substantial.
The difference may be due to Vincia's different treatment of initial and final state radiations \cite{Ritzmann:2012ca,Fischer:2016vfv}.
However, a detailed discussion of this difference falls beyond the scope of the paper.

Finally, we compare the information content in the AM and ParT for VIN-involved classification problems, using the odds ratio $\mathrm{OR}^{\I/\II}$.
The classification performance metrics for $\mathrm{OR}^{\I/\II}$ are shown in \tabref{Table:vin_marg}.
Once again, the AUCs and rejection rates $R_{X\%}$ are close to no difference results, considering the large uncertainties associated with ParT, as shown in \tabref{Table:marginbootParT}.

\begin{table}
\begin{center}
\begin{tabular}{lccccc}
\toprule
Model & 
$\VT$ vs.~$\VQ$ & 
$\PQ$ vs.~$\VQ$ & 
$\VQ$ vs.~$\HQ$ & 
$\PT$ vs.~$\VT$ & 
$\VT$ vs.~$\HT$ 
\\
\midrule
ParT-best (as $\I$) & 
0.959 &
\bf 0.623 &
\bf 0.652 &
0.604 &
\bf 0.658
\\
ParT-2nd (as $\II$) & 
0.959 &
0.623 &
0.651 &
0.602 &
0.658
\\
AM (as $\II$) & 
\bf 0.960 &
0.623
&
0.650 &
\bf 0.610 &
0.651
\\
\bottomrule
\end{tabular}
\end{center}
\caption{
AUC of the AM and ParT in classifications involving VIN-generated datasets $\VT$ and $\VQ$.
The best result for each classification problem is highlighted in boldface.
However, the differences in AUC values between AM and ParT are not statistically significant.
}
\label{tab:vinauc}
\end{table}

\begin{table}
\begin{center}
\begin{tabular}{lccccc}
\toprule
Model & 
$\VT$ vs.~$\VQ$ & 
$\PQ$ vs.~$\VQ$ & 
$\VQ$ vs.~$\HQ$ & 
$\PT$ vs.~$\VT$ & 
$\VT$ vs.~$\HT$ 
\\
\midrule
ParT-best (as $\I$) & 
186$(\pm4)$&
\bf 3.08$(\pm0.01)$ &
\bf 3.43$(\pm0.01)$ &
2.77$(\pm0.01)$  &
\bf 3.53$(\pm0.01)$
\\
ParT-2nd (as $\II$) & 
184$(\pm4)$ &
3.08$(\pm0.01)$ &
3.41$(\pm0.01)$ &
2.76$(\pm0.01)$ &
3.52$(\pm0.01)$
\\
AM (as $\II$) & 
\bf 192$(\pm4)$ &
3.07$(\pm0.01)$  &
3.40$(\pm0.01)$ &
\bf 2.84$(\pm0.01)$ &
3.42$(\pm0.01)$
\\
\bottomrule
\end{tabular}
\end{center}
\caption{Rejection rate $R_{50\%}$ of the AM and ParT in classifications involving VIN-generated datasets $\VT$ and $\VQ$.}
\label{tab:vinrejectionI}
\end{table}

\begin{table}
\begin{center}
\begin{tabular}{lccccc}
\toprule
Model & 
$\VT$ vs.~$\VQ$ & 
$\PQ$ vs.~$\VQ$ & 
$\VQ$ vs.~$\HQ$ & 
$\PT$ vs.~$\VT$ & 
$\VT$ vs.~$\HT$ 
\\
\midrule
ParT-best (as $\I$) & 
807$(\pm40)$ &
6.20$(\pm0.02)$ &
\bf 6.93$(\pm0.03)$ &
5.30$(\pm0.02)$ &
\bf 7.31$(\pm0.03)$
\\
ParT-2nd (as $\II$) & 
807$(\pm40)$ &
6.20$(\pm0.02)$ &
6.89$(\pm0.03)$&
5.26$(\pm0.02)$ &
7.28$(\pm0.03)$
\\
AM (as $\II$) & 
\bf 835$(\pm42)$ &
\bf 6.27$(\pm0.03)$ &
6.84$(\pm0.03)$ &
\bf 5.45$(\pm0.02)$ &
7.03$(\pm0.03)$
\\
\bottomrule
\end{tabular}
\end{center}
\caption{Rejection rate $R_{30\%}$ of the AM and ParT in classifications involving VIN-generated datasets $\VT$ and $\VQ$. }
\label{tab:vinrejectionII}
\end{table}

\begin{table}[t!]
    \centering
    \begin{tabular}{ccccccc}
\toprule
        $\II$& metric& 
$\VT$ vs.~$\VQ$ & 
$\PQ$ vs.~$\VQ$ & 
$\VQ$ vs.~$\HQ$ & 
$\PT$ vs.~$\VT$ & 
$\VT$ vs.~$\HT$   \\
\midrule
ParT-2nd 
& AUC & 0.5319 & 0.4965 & 0.4968 & 0.5043 & 0.4780 \\
&$R_{50\%}$ &  2.25 & 1.99 & 1.97 & 2.02 & 1.88 \\
&$R_{30\%}$ & 4.26 & 3.49 & 3.15 & 3.34 & 3.03 \\
\midrule
AM 
& AUC & 0.5104 & 0.5214 & 0.5150 & 0.4901 & 0.5293 \\
&$R_{50\%}$ & 2.09 & 2.15 & 2.09 & 1.95 & 2.17 \\
&$R_{30\%}$ & 4.07 & 3.49 & 3.39 & 3.13 & 3.67 \\
\bottomrule
\end{tabular}
\caption{ 
Classification performance metrics for the odds ratio $\mathrm{OR}^{\I/\II}$, where $\I$ is ParT-best and $\II$ is AM and ParT-2nd. 
An AUC value closer to 0.5 and rejection rates $R_{X\%}$ closer to $1/X\%$ indicates that the two classifiers have similar information contents.
}
\label{Table:vin_marg}
\end{table}

\section{Maximum Dilation Scale Choice in Morphology Analysis Module}
\label{sec:R-dep}

The morphology analysis module in \secref{subsec:pIV} has a hyperparameter: maximum dilation scale $R_\mathrm{max}$.
In this appendix, we display the influence on classifier performance under the change of this scale parameter.

The choice of this parameter may vary the performance not only in jet tagging but also in simulation comparison since the angular distribution of parton shower pattern depends on its modeling.
The subsequent emissions in parton showers with angular ordering of HW and $p_T$-ordering of PY are different \cite{Webber:2010vz,Bhattacherjee:2015psa}, and the influence can also be observed by two-point energy correlation $S_2$ based analysis \cite{Chakraborty:2019imr}.
On top of this, the Minkowski functionals capture the geometric features of parton shower in angular coordinate $(\eta,\phi)$, which are more sensitive to soft physics compared to the energy correlations.

\begin{figure}
    \centering
    \includegraphics[width=0.45\textwidth]{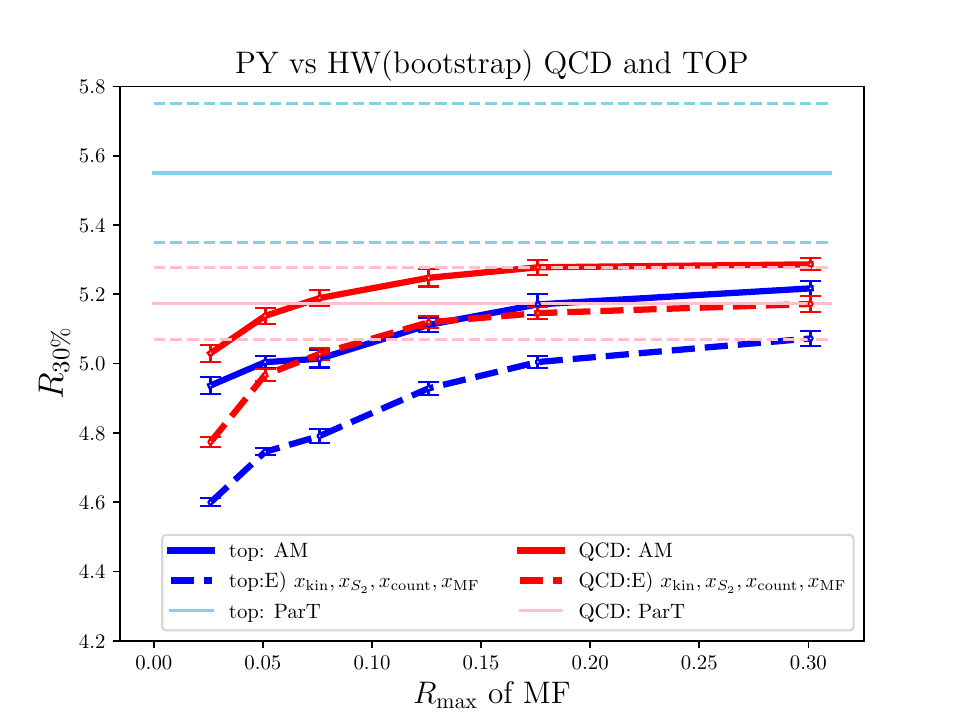}
    \includegraphics[width=0.45\textwidth]{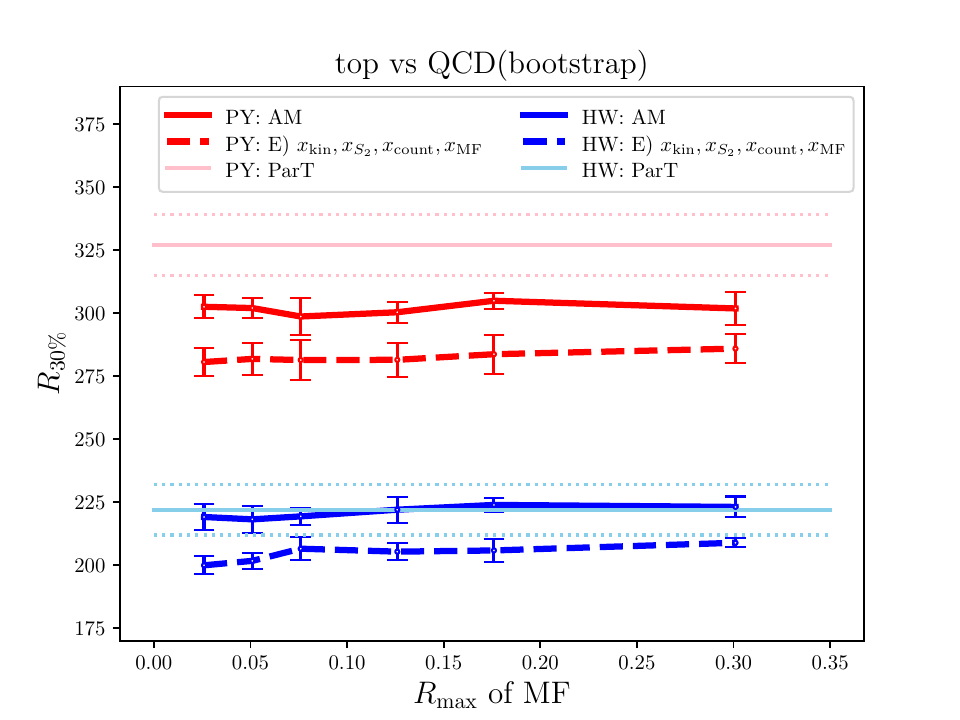} 
    \caption{
    Left: 
    The $R_\mathrm{max}$ dependence of $R_{30\%}$ of the $\PT$ vs.~$\HT$ classification (blue lines) and $\PQ$ and $\HQ$ classification (red lines). 
    The solid lines are for the AM, and the dashed lines are for the AM-PIP E) $x_\kin$, $x_{S_2}$, $x_\MF$ + $x_\rcount$. 
    The $R_{30\%}$ of the ParT model is also shown in horizontal lines. 
    All errors are estimated by bootstrapping. 
    Right: The same plot for $\PT (\HT)$ vs.~$\PQ (\HQ)$ classification.  
    }
    \label{fig:bs_Rdep}
\end{figure}

In \figref{fig:bs_Rdep}, we show the rejection rates $R_{30\%}$ of the AM for different choices of $R_\mathrm{max}$. 
The left plots show the PY vs.~HW classification result for top and QCD jets.
Each point in the plot is the average and standard deviation of $R_{30\%}$ for the different bootstrap datasets. 
We can see that too small $R_\mathrm{max}$ results in degraded performance since topological and geometrical features prominent at a larger dilation scale are not counted yet, but as $R_\mathrm{max}$ increases, the rejection rate approaches that of ParT and starts to saturate from $R_\mathrm{max} = 0.3$

In the same plot, we also show in dashed line the $R_\mathrm{max}$ dependence of AM-PIP E) taking the following inputs: $x_{\kin}$, $x_{S_2}$, $x_\rcount$ and $x_{\MF}$.
AM shows a weaker dependence on $R_\mathrm{max}$ compared to AM-PIP E).
This is because $x_{\MF}$ and the subjet constituent multiplicies in $x_{\subj_{ex}}$ for $R=0.1, 0.2,$ and $0.3$ are partially correlated.  
The residual $R_\mathrm{max}$ dependence of  $R_{30\%}$ of AM comes from the jet constituents outside the leading subjets. 
The variances of rejection rates of AM under the scaling of $R_\mathrm{max}$ is around 5\% for AM  and 10\% AM-PIP E) within the range [0.025, 0.3], and they are also statistically significant.

As discussed in the main text, the estimated error of $R_{30\%}$ for the ParT model is significantly larger than that for AM(-PIP) models.  
The solid and dashed horizontal lines show $R_{30\%}$ of ParT and its $1\sigma$ deviation estimated by bootstrapping. 
The deviation is almost as large as the changes of $R_{30\%}$ of AM-PIP from $R_\mathrm{max}=0.025$ to 0.3.  
This illustrates that AM is beneficial in interpreting the classifiers, such as precision analysis on the influence of input features on the classification performance.

In the right plot, we show the results of $\PT$ vs.~$\PQ$ and $\HT$ vs.~$\HQ$ classification. 
Unlike the top jet tagging problems, $\PQ$ vs.~$\HQ$ and  $\PT$ vs.~$\HT$, we do not see significant $R_\mathrm{rmax}$ dependences. 
This is expected because the difference of the hard substructure is the dominant part of the classification, and the parton shower emissions and the hadronization model are common between $\PT$ and $\PQ$ or $\HT$ vs.~$\HQ$.

\section{Comparing Computational Resources utilized by AM and ParT}\label{sec:scomp}

In this appendix, we describe the computation resources utilized by AM and ParT.
During training with a batch size of 1,000, AM requires less than 1 GB of GPU memory, while ParT consumes approximately 14 GB.
Additionally, the training process of AM is significantly faster.
When training AM on an NVIDIA GeForce 1080 Ti GPU (FP32 performance: 11.3~TFLOPS), the training session takes 30 minutes with an overall GPU utilization of 35\%.
In comparison, training of ParT on an NVIDIA RTX 6000 GPU (FP32 performance: 38.7~TFLOPS) also takes 30 minutes, but with a GPU utilization of 95\%. 
The ratio of the number of floating-point operations, calculated as (FP32 performance)~$\times$~(GPU utilization)~$\times$~(training time), between ParT and AM is approximately 18.3 to 2.0.
This indicates that AM requires a significantly smaller number of GPU operations than ParT.

\section{Proof of Classifier Interpretation of Odds Ratio}
\label{app:proof_s_marg}

In this appendix, we derive that $s^{\I/\II}$ is a classifier that relies solely on the features exclusive to the more expressive model. 
Let $\I$ be a more expressive model than $\II$ without loss of generality. 
We denote the common feature used in both $\I$ and $\II$ as $x_2$, and the exclusive feature as $x_1$.
The classifier predictions $s^\I$ and $s^\II$ are regressing the posteriors for class 1,
\begin{equation}
    s^\I(x_1, x_2) = p(y=1\,|\,x_1, x_2), \quad s^\II(x_2) = p(y=1\,|\,x_2).
\end{equation}
The odds ratio is then written as a posterior odds ratio,
\begin{equation}
    \mathrm{OR}^{\I/\II}(x_1, x_2) = \frac{p(y=1\,|\,x_1, x_2)}{p(y=0\,|\,x_1, x_2)} \cdot \frac{p(y=0\,|\, x_2)}{p(y=1\,|\, x_2)}.
\end{equation}
By expanding the conditional probabilities, we can simplify the odds ratio to likelihood ratio between class labels given input $x_1$, conditioned on $x_2$,
\begin{equation}
    \mathrm{OR}^{\I/\II}(x_1,x_2) = \frac{p(x_1 \,|\, y=1, x_2)}{p(x_1 \,|\, y=0, x_2)}.
\end{equation}
From this point, let us regard $x_2$ in the above expression as a conditioning parameter of distribution rather than a random variable. 
We denote probability $p(\cdots, x_2)$ conditioned on $x_2$ as $p_{x_2}(\cdots)$.
The function $s^{\I/\II}$ is then written as follows:
\begin{equation}
    s^{\I/\II}(x_1,x_2)
    = 
    \frac
    {\mathrm{OR}^{\I/\II}(x_1,x_2)}
    {1+\mathrm{OR}^{\I/\II}(x_1,x_2)}
    =
    \frac{p_{x_2}(x_1 \,|\, y=1)}{p_{x_2}(x_1 \,|\, y=1) + p_{x_2}(x_1 \,|\, y=0)} 
\end{equation}
The expression on the right-hand side is the posterior probability $p_{x_2}(y=1 | x_1)$ with a class prior $p_{x_2}(y) = 0.5$.
Therefore, $s^{\I/\II}$ can be interpreted as a classifier using only the exclusive feature $x_1$, given a fixed common feature $x_2$.

\bibliographystyle{JHEP}
\bibliography{HLF_Anatomy_TopJets}

\end{document}